\documentclass[preprint,aps,floatfix,superscriptaddress,pre]{revtex4}
\usepackage{amsmath, amsthm, amssymb, graphicx}
\usepackage{color}
\usepackage{multirow}
\usepackage{ulem}
\usepackage[caption=false]{subfig}
\usepackage{adjustbox}

\input epsf.sty
\begin{document}

\title{Perturbing cyclic predator-prey systems: how a six-species coarsening system with non-trivial in-domain dynamics
responds to sudden changes }
\author{Shadisadat Esmaeili}
\affiliation{Department of Physics, Virginia Tech, Blacksburg, VA 24061-0435, USA}
\affiliation{Center for Soft Matter and Biological Physics, Virginia Tech, Blacksburg, VA 24061-0435, USA}
\author{Barton L. Brown}
\affiliation{Department of Physics, Virginia Tech, Blacksburg, VA 24061-0435, USA}
\affiliation{Center for Soft Matter and Biological Physics, Virginia Tech, Blacksburg, VA 24061-0435, USA}
\author{Michel Pleimling}
\affiliation{Department of Physics, Virginia Tech, Blacksburg, VA 24061-0435, USA}
\affiliation{Center for Soft Matter and Biological Physics, Virginia Tech, Blacksburg, VA 24061-0435, USA}
\affiliation{Academy of Integrated Science, Virginia Tech, Blacksburg, VA 24061-0405, USA}
\date{\today}

\begin{abstract}
Cyclic predator-prey systems have been shown to give rise to rich, and novel, space-time patterns, as for example
coarsening domains with non-trivial in-domain dynamics. In this work we study numerically
the responses of a cyclic six-species model, characterized by the formation of spirals inside coarsening domains,
to two different types of perturbations: changing the values of the predation and reproduction rates as well as
changing the interaction scheme. For both protocols we monitor the time evolution of the system after the onset of the perturbation
through the measurement of dynamical correlation functions and time-dependent densities of empty sites. 
In this way we gain insights into the
complex responses to different perturbations in a system where spirals, which are due to 
the formation of cyclic alliances,
dominate the dynamics inside the coarsening domains.

\end{abstract}

\maketitle

\section{Introduction}
Issues around biodiversity and species extinction in ecological networks \cite{May74,Smith74,Sole06} have yielded an increased interest
among statistical physicists \cite{Szabo07,Frey10}, due to the many novel, and often unexpected, features that emerge when
going beyond the mean-field treatment of the simplest predator-prey models. Already simple modifications, like adding stochastic
effects \cite{McK95} and/or a spatial environment \cite{Mob06,Mob07}
to the standard Lotka-Volterra model, can markedly change species coexistence and extinction. Also, going beyond a simple predator-prey
relationship by allowing for more than two species yields new scenarios that have received much attention in recent years
(see \cite{Frey10,Szolnoki14,Dobramysl18} for recent reviews). In this context spatial cyclic predator-prey models that allow for different
interaction schemes have attracted special attention lately 
\cite{Roman13,Mowlaei14,Roman16,Brown17,Avelino12a,Avelino12b,Avelino14a,Avelino14b,Avelino17a,Avelino17b,Labavic16}
due to the large variety of intriguing space-time patterns as for example domain coarsening with non-trivial
in-domain dynamics in the form of spirals \cite{Brown17}.

The stability of an ecological system, i.e. the tendency of the system to return to its
steady state after a perturbation, has been the subject of many studies in the last forty years \cite{May74,May72,Mcc00}.
Many of these studies focused on the relationship between stability and the underlying network
structure \cite{May74,Pas06,Bas10}, which yielded important insights into the generic stability properties
of networks with predator-prey
interactions.
Cyclic games provide interesting situations, where on the level of the rate equations
a variety of situations can be realized (heteroclinic cycles, neutrally stable closed orbits, limit cycles),
depending on the presence or absence of particle number conservation and the inclusion or exclusion
of mutations \cite{Szolnoki14,Dobramysl18}. Of course, these different dynamical situations yield different behaviors
under perturbations. Further complications appear when considering stochastic and spatial effects
in the responses of finite populations to perturbations.

In this context one important aspect has not received
adequate attention in the past: how do these systems approach the steady state
after the perturbation? Different scenarios are possible, as for example: (1) the perturbation is finite in time
and the system evolves back into the same steady state as before the perturbation; (2)
the system is stable and returns to its steady state even when the (weak) perturbation persists; (3) the system is unstable
or the perturbation is a massive one, and the system ends up in a different steady state.

In this work we consider six-species cyclic predator-prey models that we subject to different perturbations.
Our goal is to gain an understanding of how complex systems like these respond to changes and of how
they approach the (original or new) steady state. Depending on the interaction scheme, the six-species cyclic predator-prey 
game yields different space-time patterns. The two interaction schemes on which we focus are characterized by the appearance of coarsening
domains with spirals forming inside the domains or by a growth process where different domain types, each containing an alliance
formed by two neutral partners, compete. We are not aware of any real-world examples described by the interaction schemes
discussed in this paper. Still, from the point of view of pattern formation far from equilibrium, our study allows us to
gain insights into mechanisms that permit systems with complex space-time patterns to adapt to different perturbations.

The different scenarios considered in the following are inspired by real-world situations. We first investigate how a system
reacts to changes to the interaction and reproduction rates, thereby mimicking the pressure put on an ecological system by
changes in environmental conditions. The second perturbation consists in changing the interaction scheme. Such a rewiring
changes the relationships between the species, and new alliances replace existing ones. Besides the situation where the system
is provided with enough time to fully adapt to the new relationships, we also consider the case of a periodic switching
between the two interaction schemes with a frequency large enough that the system can not fully adjust to the previous
change before the next change takes place. This last case can be seen as an attempt at implementing a protocol that reflects seasonal variations.

As the results discussed in the following show, the presence of spirals, which emerge due to the formation of cyclic alliances,
has a major impact on the response of a coarsening system to perturbations. Changing the predation rate while keeping the interaction
scheme unchanged yields in presence of spirals an abrupt and non-monotonous change of the correlation length and the density of empty sites. 
This surprising behavior reflects the adjustment of the spiral size inside the domains.
After this initial quick response, the system enters a second regime characterized by
an algebraic relaxation. On the other hand, changing the interaction scheme, where we switch between a scheme that favors cyclic alliances
and a scheme with neutral alliances, results in a complicated behavior of the correlation length due to the dissolution of old and the formation of
new alliances. 
Switching from cyclic alliances to neutral alliances results in a quick disordering during which neutral partners spontaneously
aggregate, yielding a shallow dip in the time-dependent correlation length. Switching from neutral alliances to cyclic alliances, on the 
other hand, results in a two-stage process as revealed by two drops of the correlation length. The first quick drop is due to the start of
predation events between previously neutral species, whereas the second, larger drop is a consequence of a large-scale rearrangement 
as existing domains are dissolved and replaced by new domains with in-domain spirals.

\section{Model}
Cyclic predator-prey models have been shown to produce complex patterns 
in space and time (see \cite{Szabo07,Frey10,Szolnoki14,Dobramysl18} and references therein). We denote by
$(N,r)$ a family of cyclic models \cite{Roman13} with $N$ species where each species
attacks $r$ other species in a cyclic manner, i.e. species $i$ preys on species $i+1$, $i+2$, $\cdots$,
$i+r$ (this has to be understood modulo $N$). The type and complexity of the emerging patterns (coarsening domains, 
spirals, or a combination of both) depend on the number of species ($N$) and the interaction scheme ($r$) involved in the game
\cite{Roman13,Mowlaei14,Roman16,Brown17,Avelino12a,Avelino12b,Avelino14a,Avelino14b,Avelino17a,Avelino17b,Labavic16}.

In most of this work, we assume a two-dimensional lattice where species interactions are limited to the four nearest neighbors.
We consider a May-Leonard-type system described by the following set of reactions taking place between nearest neighbors:
\begin{equation}
\begin{split}
s_{i} + s_{j} &\xrightarrow[]{\kappa} s_{i} + \emptyset \\
s_{i} + \emptyset &\xrightarrow[]{\kappa} s_{i} + s_{i}\\
s_{i} + X &\xrightarrow[]{\sigma} X+s_{i} 
\end{split}
\label{eq:1}
\end{equation}
where $s_{i}$ denotes an individual of species $i$, whereas $s_j$ is an individual of species $j$ preyed upon by species $i$.
As a result of the predation event, described by the first reaction in (\ref{eq:1}),
the individual $s_j$ is removed, yielding an empty lattice site indicated by $\emptyset$.
This empty site can be occupied through a reproduction process where the individual $s_{i}$ creates an offspring at a neighboring
empty site, see the second reaction in (\ref{eq:1}). 
In the last reaction of (\ref{eq:1}) $X$ stands for an individual of any species or for an empty site. This reaction
makes sure that individuals are mobile and can swap places with their neighbors or jump to a neighboring empty site.
In this paper we only consider the situation that $\kappa + \sigma =1$.

In our numerical simulations we always prepare a square system of $L \times L$ sites (for the data discussed below we have
$L=700$, but we checked that none of our conclusions depend on the system size) in a disordered initial state where every lattice site 
is either empty or occupied with the same probability $1/(N+1)$ by each species. 
For every update we first select randomly a lattice site before selecting randomly
one of its neighbors. Depending on which species (if any) occupy the two selected sites, one of the reactions (\ref{eq:1}) takes place
with a chosen rate ($\kappa$ for predation and reproduction and $\sigma$ for swapping, with $\kappa + \sigma =1$). If the two sites
are occupied by individuals from two species that prey on each other, then the individual sitting on the first selected site is considered
to be the predator. 
As usual, time is measured in Monte Carlo steps, with one Monte Carlo step corresponding to $L^2$ proposed updates.

In addition to the system on a two-dimensional lattice, we also briefly consider the well-mixed case 
without an underlying lattice where every individual can interact with every other individual. In that case only the
simultaneous predation and reproduction event (with rates $\kappa_{ij}$ that might depend on the predator and prey species $i$ and $j$)
\begin{equation}
s_{i} + s_{j} \xrightarrow[]{\kappa_{ij}} s_{i} + s_{i}
\end{equation} 
takes place.

\begin{figure} [h]
\includegraphics[width=0.275\columnwidth,clip=true]{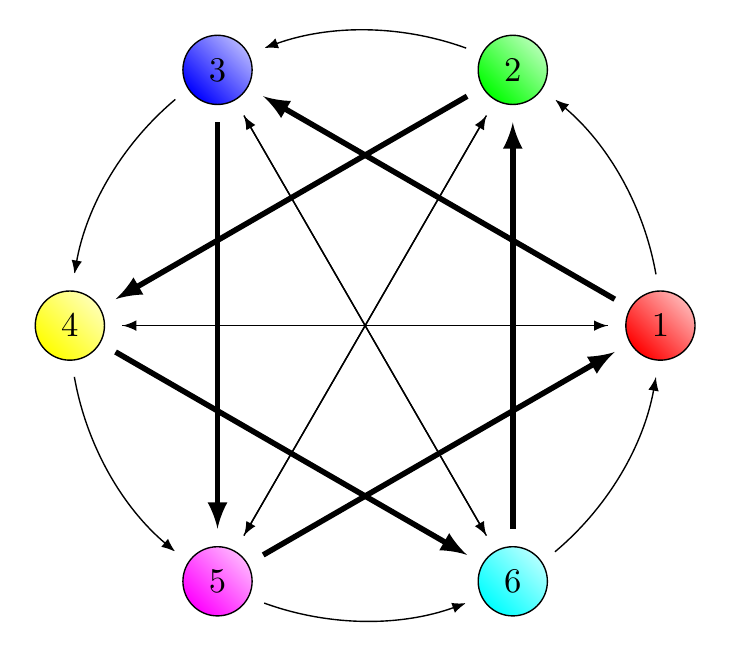}
\caption{\label{fig1} The $(6,3)$ game interaction diagram. The arrows connect predators with their preys.
In two space dimensions one observes the formation of domains composed by two different teams of cyclically
interacting species, as indicated by the bold arrows.
}
\end{figure}

\begin{figure} [h]
\centering
        \subfloat[]{
        \resizebox{0.25\textwidth}{0.225\textwidth}{%
        \trimbox{0.25cm 0.2cm 0.6cm 0.2cm}{
                \includegraphics{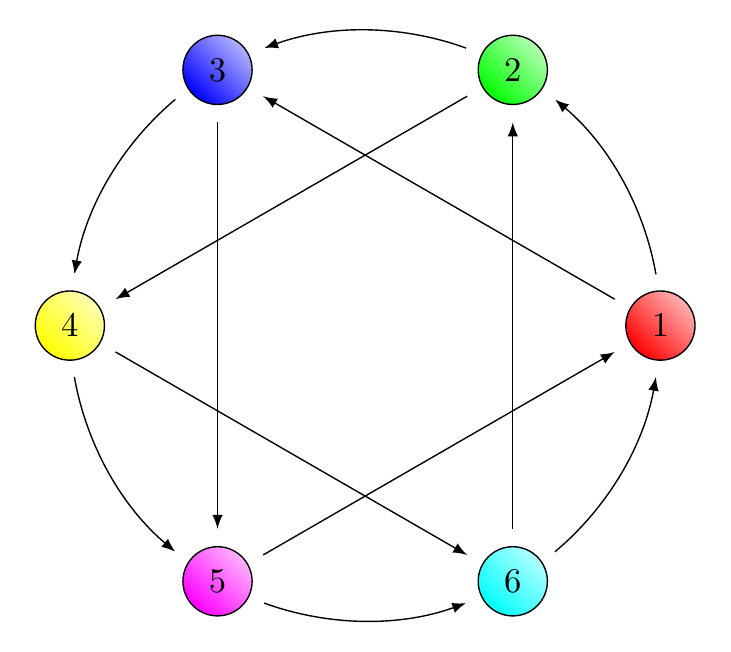}
        }
        }
        }
  \subfloat[]{%
        \resizebox{0.25\textwidth}{0.225\textwidth}{%
        \trimbox{0.25cm 0.22cm 0.56cm 0.22cm}{
                \includegraphics{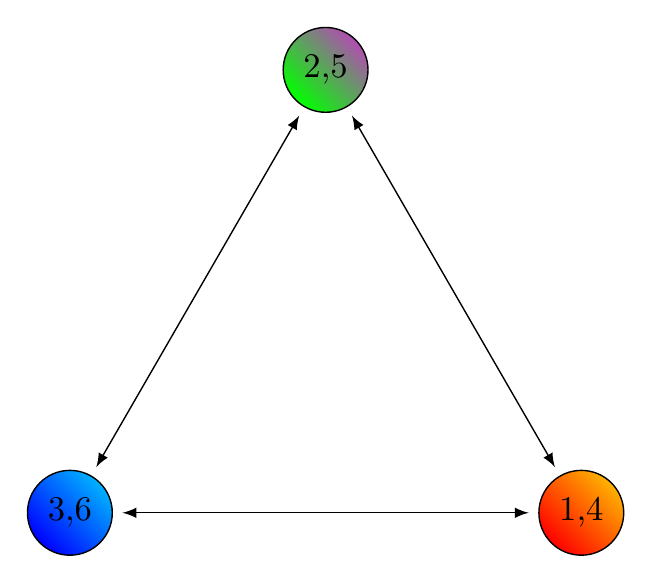}
        }
        }
        }
\caption{\label{fig2} (a) The $(6,2)$ game interaction diagram. Each species has a neutral partner where swapping is the only allowed interaction.
One therefore observes the formation of three teams composed each of two neutral partners. (b) The three different teams compete with each other.
}
\end{figure}

\begin{figure} [h]
\minipage{0.30\textwidth}
  \centering
  \includegraphics[width=\linewidth]{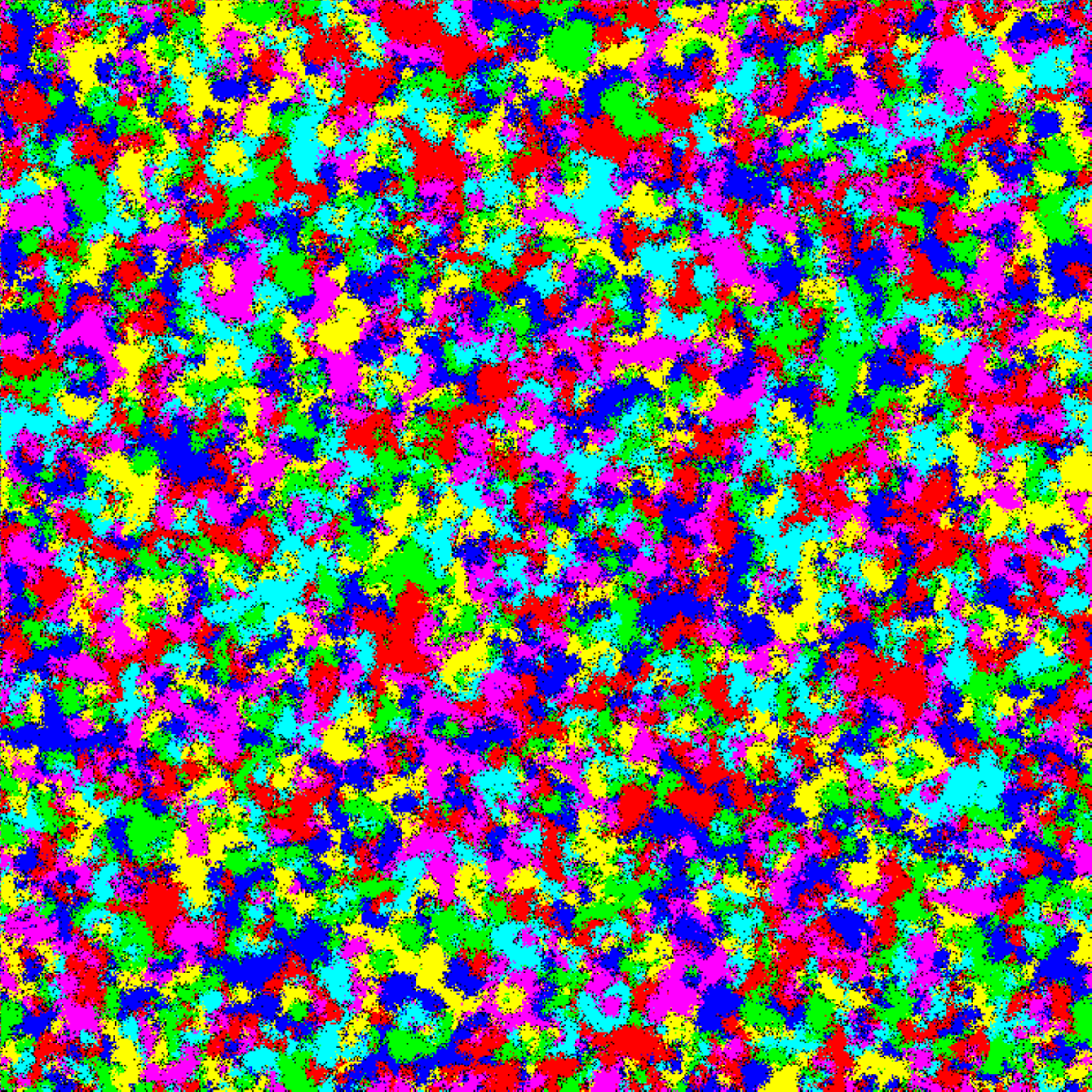}\\
  t=100
\endminipage
\hspace*{0.05\textwidth}
\minipage{0.30\textwidth}
  \centering
  \includegraphics[width=\linewidth]{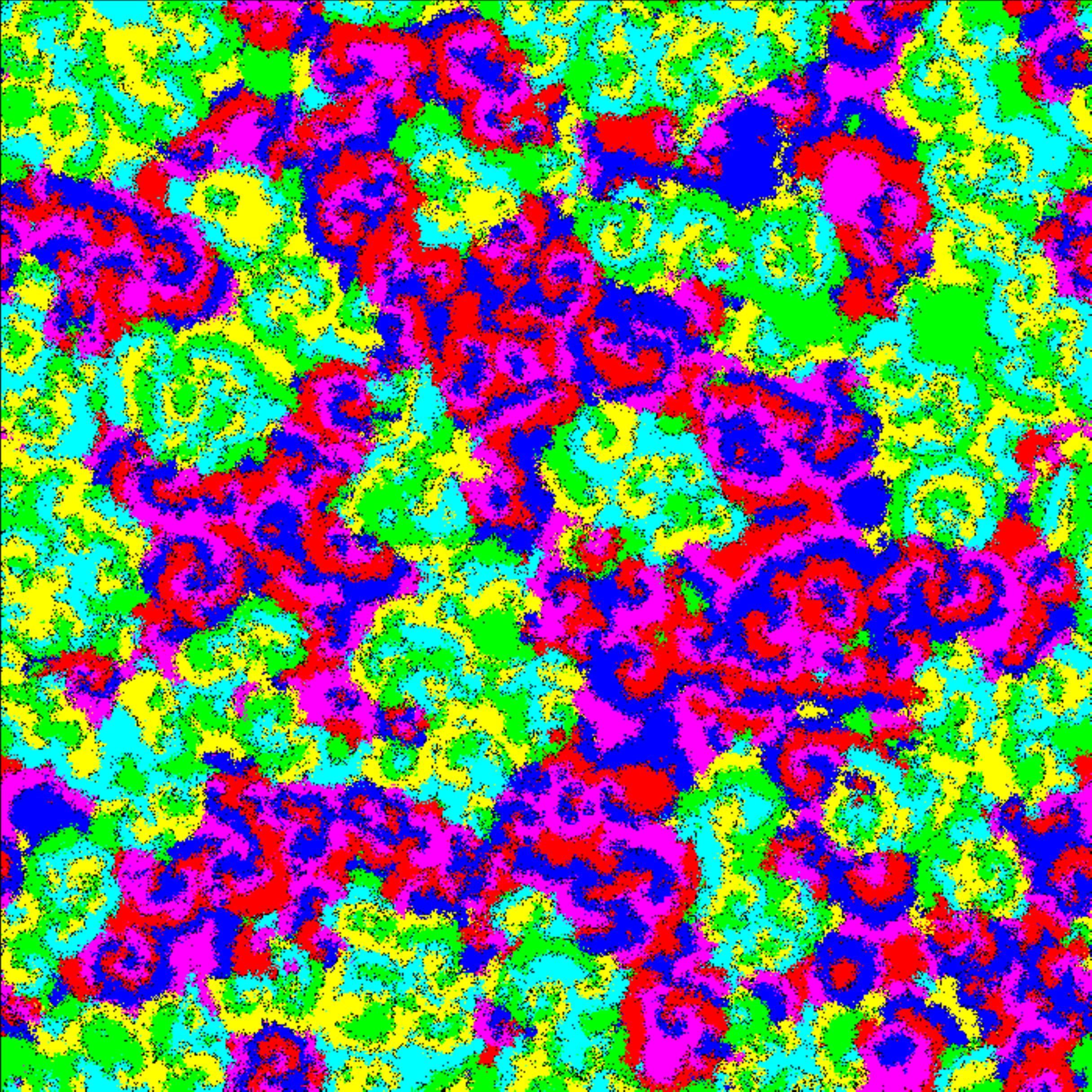}\\
  t=500
\endminipage\\
\caption{\label{fig3} When starting from a disordered initial state the $(6,3)$ game on a lattice results in the formation
of domains where each domain is occupied by a team of three species playing a $(3,1)$ rock-paper-scissors game. The 
snapshots have been taken at times $t=100$ and $t=500$ after the initial preparation.
The rates have been set to $\kappa=\sigma = 0.5$.
}
\end{figure}

\begin{figure} [h]
\minipage{0.30\textwidth}
  \centering
  \includegraphics[width=\linewidth]{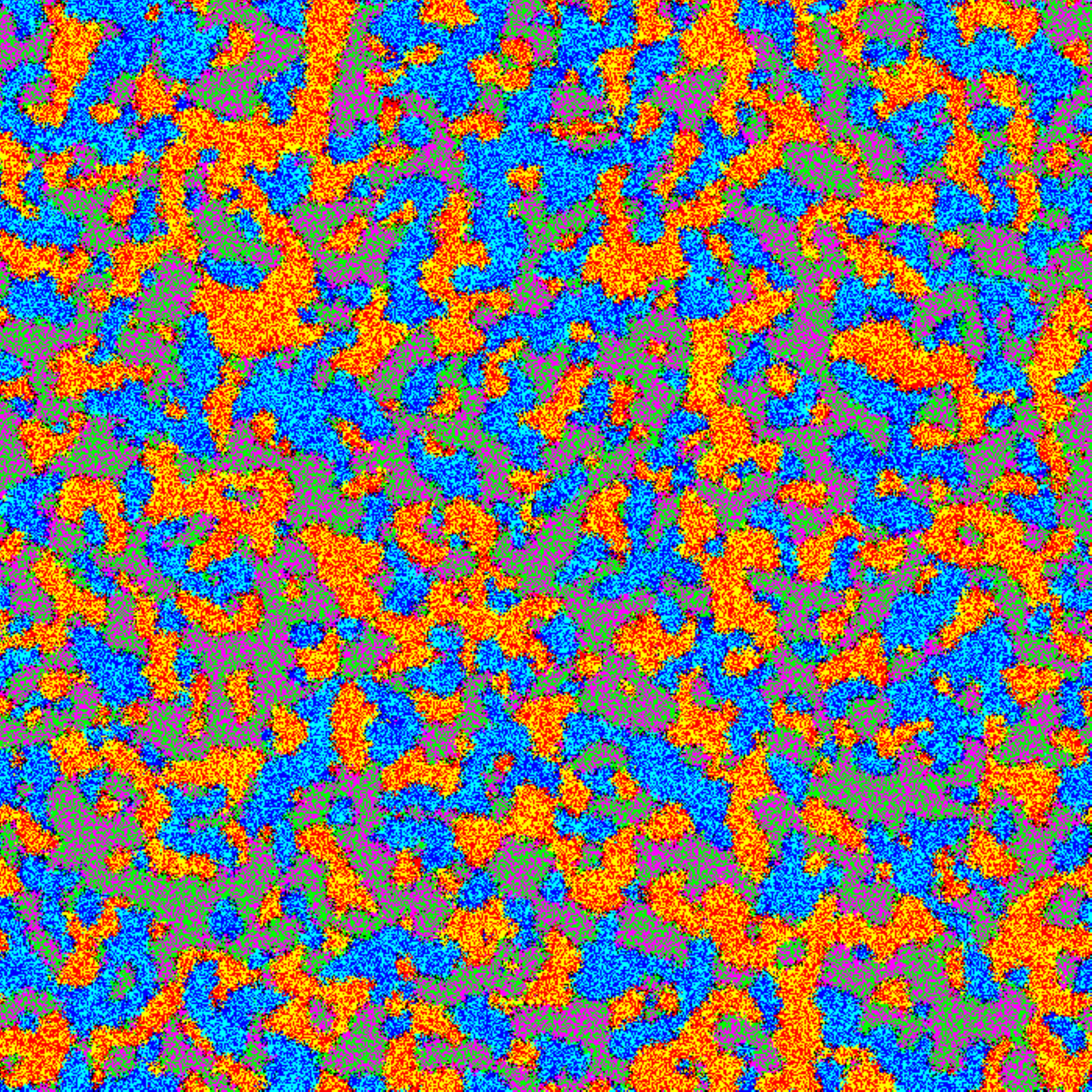}\\
  t=100
\endminipage
\hspace*{0.05\textwidth}
\minipage{0.30\textwidth}
  \centering
  \includegraphics[width=\linewidth]{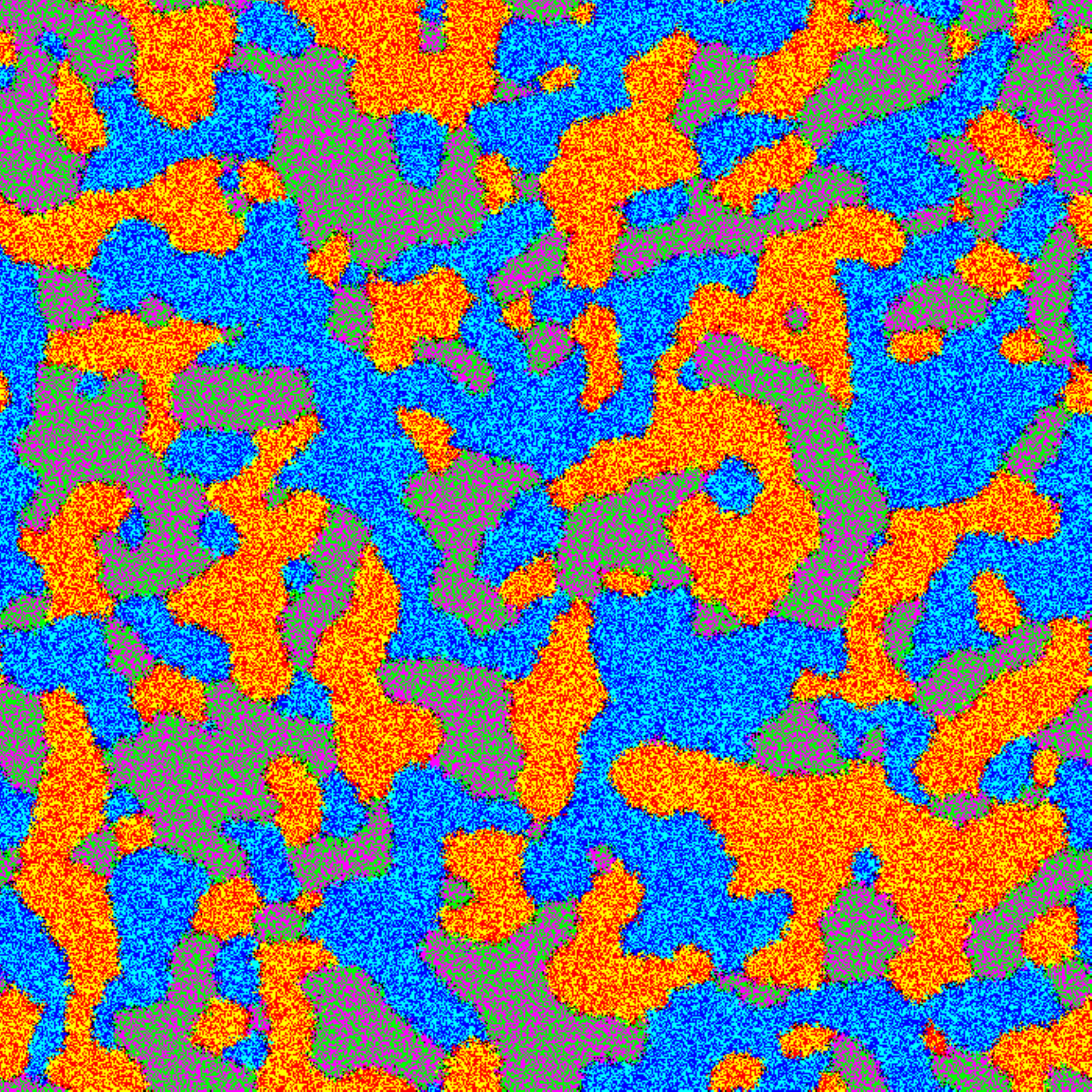}\\
  t=500
\endminipage\\
\caption{\label{fig4} The $(6,2)$ game is characterized by the formation of three teams of two neutral partners, yielding
a coarsening process involving three different domain types. The
snapshots have been taken at times $t=100$ and $t=500$ after the initial preparation.
The rates have been set to $\kappa=\sigma = 0.5$.
}
\end{figure}

Our focus is on the $(6,3)$ and $(6,2)$ games, two six-species games where species interact
following the schemes shown in Figs. \ref{fig1} and \ref{fig2}. In the $(6,3)$ game the species organize
themselves into two different teams \cite{Roman13,Brown17} as indicated by the bold arrows in Fig. \ref{fig1}. These two teams
compete with each other while playing an in-team cyclic game. This leads to the formation of spirals within 
coarsening domains, as illustrated in the snapshots shown in Fig. \ref{fig3}. Changing the scheme from $(6,3)$
to $(6,2)$, i.e. going from three prey for each species to only two, results in a fundamental change of the emerging space-time
patterns. Indeed, for the $(6,2)$ game we have three pairs of neutral partners as, e.g., species 1 and 4. It is advantageous for these
neutral partners to agglomerate as this allows each species to be protected by their partner. Consequently, see Fig. \ref{fig4}, 
one observes the formation of three types of domains, each occupied by one team, that undergo a coarsening process.

The spatio-temporal properties of the $(6,3)$ game in two space dimensions and $\kappa = 0.5$ have been the subject of a recent study
\cite{Brown17}, so that we will in the following only summarize the main properties in as much as they are needed for
the present investigation. As shown in \cite{Brown17} the formation of spirals within coarsening domains has a non-trivial impact
on both domain growth and interface fluctuations. Indeed, for $\kappa = 0.5$ the dynamic correlation length increases as $t^{0.43}$, whereas 
for standard curvature-driven coarsening this length is expected to increase with an exponent 1/2, as it is the case for the 
standard Ising model. Similarly, aging exponents are found to differ from those encountered for the two-dimensional Ising
model \cite{Henkel10}. Finally, the growth and roughness exponents governing the time-dependence of the interface width 
have values $\beta= 0.43$ and $\alpha = 0.15$ that are not identical to those of the standard Edwards-Wilkinson exponents 
$\beta_{EW} = 1/4$ and $\alpha_{EW} = 1/2$ \cite{Edwards82}. These non-standard values for the different
exponents are a consequence of the spiral wave fronts that induce large-scale coherent fluctuations at the interfaces
separating different domains \cite{Brown17}.

Inside the domains that form for the (6,2) interaction scheme individuals belonging to the two neutral species simply move around, either
by exchanging places or by jumping into an empty site. This only changes at the interface between domains where predators and preys
interact. The situation is therefore similar to that encountered for the $(4,1)$ game \cite{Roman12} or the $(2,1)$ game \cite{Brown17},
with the exception that in the present case we end up with three competing domain types as compared to the two domain types 
encountered for $(4,1)$ and $(2,1)$. For all these cases one should have the same dynamic and roughness exponents than for the
two-dimensional Ising model quenched below the critical temperature. This has been verified previously for $(4,1)$ and $(2,1)$.
For $(6,2)$ we checked that the dynamic exponent governing the coarsening process indeed takes on the value $z=2$, yielding
a dynamical length that increases as $t^{1/2}$.

In order to understand how a coarsening system with non-trivial in-domain dynamics reacts to perturbations, we discuss in the
remainder of this paper how the (6,3) game adjusts to two specific perturbations, namely a sudden change of the values of the rates
as well as a sudden change of the interaction scheme.

In our simulations we register the time evolution of the number densities for each species as well as of the density of empty sites.
Following \cite{Brown17} we distinguish two different empty site populations: (1) the empty sites created within domains due to
predator-prey interactions between team members and (2) the empty sites created at the domain boundaries resulting from the interactions
between the teams. Important insights into the space-time properties of our system are provided by space-time correlation functions
and the time-dependent lengths extracted from them. For our six-species system the standard space-time correlation function is given by
\begin{equation} \label{Cstandard}
C(t,r) = \sum\limits_{i=1}^6 \left[ \left< n_i(\vec{r},t) n_i(0,t) \right> - \left< n_i(\vec{r},t) \right> \left< n_i(0,t) \right> \right]~,
\end{equation}
where $r= \left| \vec{r} \, \right|$, whereas $n_i(\vec{r},t)$ is 1 if at time $t$ an individual from species $i$ occupies site $\vec{r}$ and zero otherwise. 
$\left< \cdots \right>$ indicates an ensemble average over noise histories and initial conditions. As we observe the formations of teams, we also
consider a team-team space-time correlation function where all species forming one team are viewed as belonging to the same type:
\begin{equation} \label{CT}
C_T(t,r) = \sum\limits_{j=1}^T \left[ \left< m_j(\vec{r},t) m_j(0,t) \right> - \left< m_j(\vec{r},t) \right> \left< m_j(0,t) \right> \right]~,
\end{equation}
where $T$ is the number of teams (two for the (6,3) game, for example) and $m_j(\vec{r},t) =1$ if the site $\vec{r}$ is occupied at time $t$
by an individual belonging to one of the species forming team $j$ and zero otherwise. The function $C_T(t,r)$ therefore focuses on the space-time
patterns emerging due to the competition of the species without revealing correlations that result from the interactions between team members.

In order to extract a time-dependent length $L(t)$ from the correlation function $C(t,r)$, we determine the distance at which the normalized
space-time correlation function takes on a specific value $C_0$:
\begin{equation}
C(t,L(t))/C(t,0) = C_0~.
\end{equation}
Similarly we also obtain a correlation length $L_T(t)$ from the team-team space-time correlation function $C_T(t,r)$.

We note that other methods for extracting a length from the space-time correlation function have been proposed. Common approaches
include collapsing curves for different $t$ by scaling $r/L(t)$ or using the method of integral
estimators \cite{Belleti08,Park12}. Both methods assume that the scaling function is a function of $r/L(t)$ only. However, as we have shown in
\cite{Brown17}, see Fig. 4 in that paper, this is not the case for systems with spirals inside coarsening domains.

\section{Changing the rates}
Changes in the environment may result in changes to the efficiency of a predator and/or changes to the birth rate of a species.
In the model considered here, both predation rate and reproduction rate are identical and given by $\kappa$. The mobility
$\sigma$, which can be realized through
jumps to empty sites or swaps of the positions of two neighboring individuals, is related to $\kappa$ as we fix $\kappa + \sigma = 1$.
Therefore, if the individuals are very mobile, they are less efficient hunters and give birth to off-springs at a lower rate.
Similarly, low mobility implies efficient predation and high birth rate.

In the following we consider cases where $\kappa$ is changed from an initial value $\kappa_i$ to a final value $\kappa_f$. We focus
on the values $\kappa=0.3$, 0.5 and 0.7, and we checked that a qualitative similar behavior is encountered for other
values of $\kappa$.

\begin{figure} [h]
\includegraphics[width=0.45\columnwidth,clip=true]{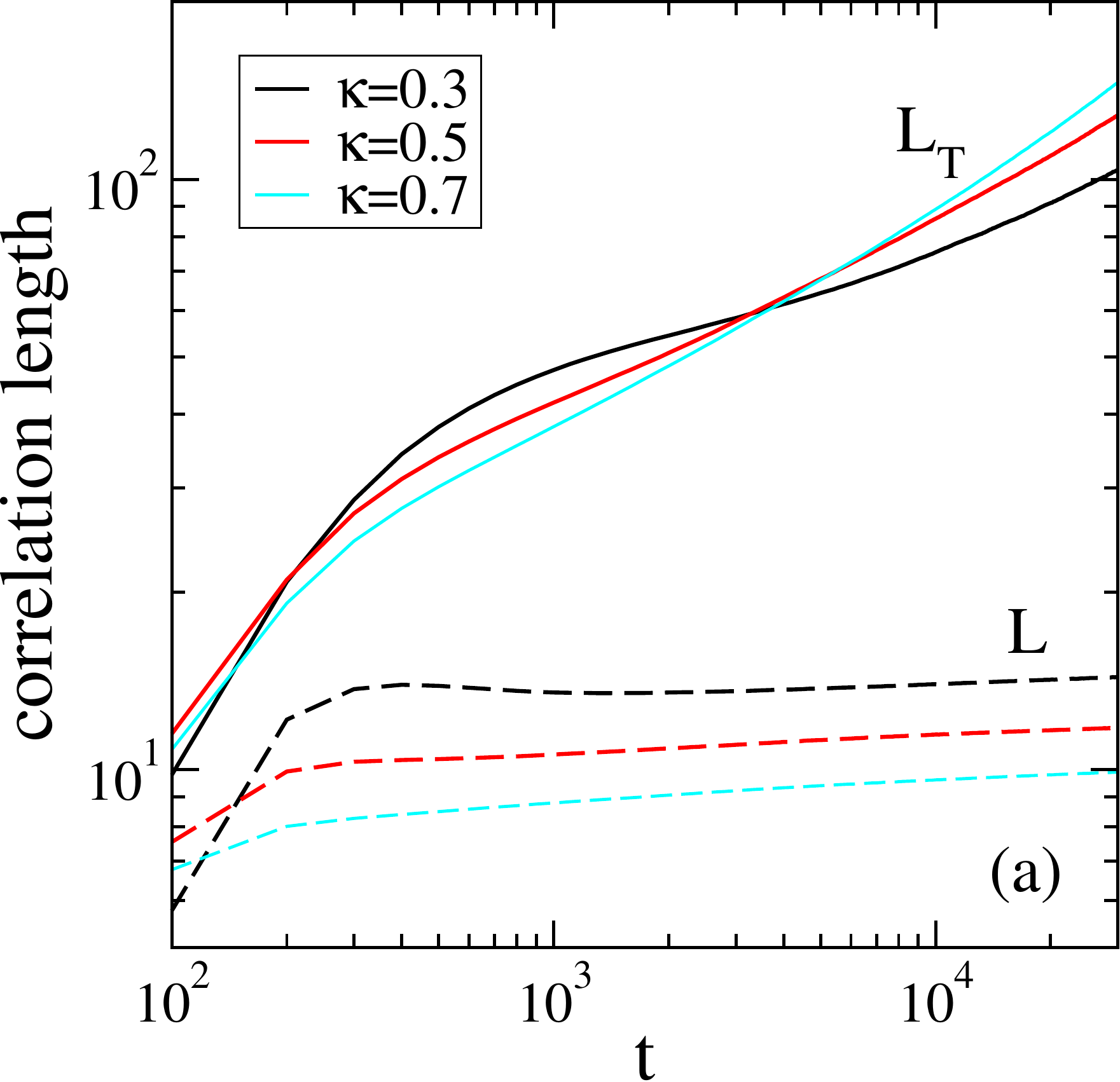}\quad
\includegraphics[width=0.45\columnwidth,clip=true]{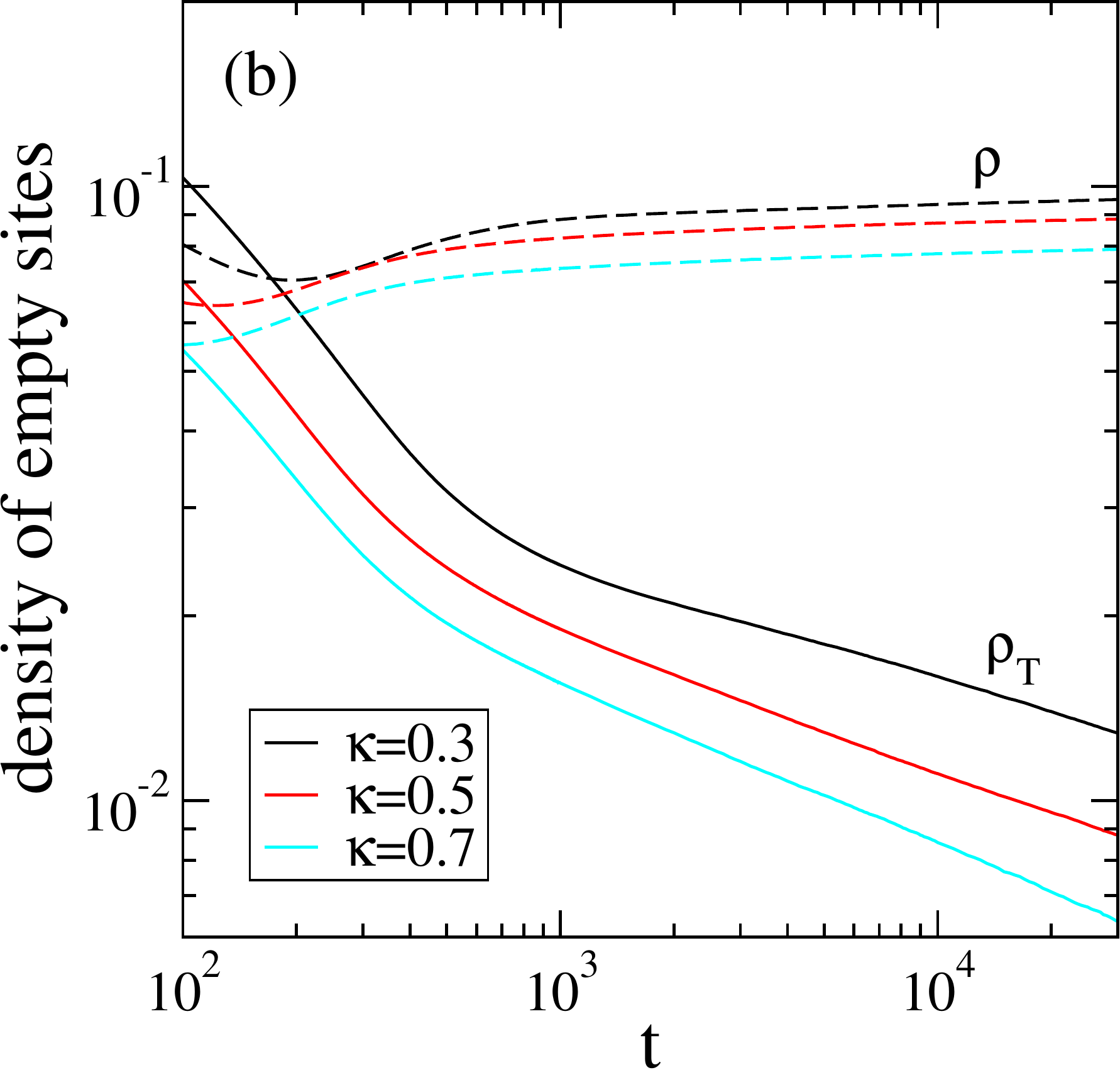}
\caption{\label{fig5} Time-dependent correlation lengths (a) and densities of empty sites (b)
for systems composed of $700 \times 700$ spins evolving at constant rates $\kappa = 0.3$, 0.5, and 0.7.
In (a) the dashed lines are obtained from the intersections of the standard space-time correlation
function $C(t,L(t))$ with the horizontal line $C_0 = 0.2$, whereas the full lines result from determining the
lengths at which the team-team space-time correlation function $C_T(t,L_T(t))=0.1$. In (b) the densities of empty sites
$\rho_T(t)$ that follow from reactions between the two teams are shown as full lines, whereas the dashed lines indicate the
densities of empty sites $\rho(t)$ produced in reactions involving members from the same team. The data follow from averaging over at
least 4000 independent runs.
}
\end{figure}

Before changing the rate during the coarsening process, we first need to understand the unperturbed systems that evolve with constant
rates. Fig. \ref{fig5} shows the time evolution of different correlation lengths and empty site
densities for systems with constant rates $\kappa = 0.3$ 0.5, and 0.7. We can distinguish two different regimes:
an early time regime where the formation of domains coincides with the establishment of spirals and a late time coarsening
regime where multiple spirals fit inside the growing domains. As shown in
Fig. \ref{fig5}a, the time-dependent lengths obtained from the standard space-time correlation function (\ref{Cstandard}) 
with $C_0=0.2$ ($L(t)$, dashed lines) and
the team-team correlation function (\ref{CT}) with $C_0 = 0.1$ ($L_T(t)$, full lines) behave very differently. The length $L(t)$, which 
for larger values of $C_0$ provides a 
proxy for the width of the spirals, tends toward a plateau once the
spirals have formed. It follows from the data shown in Fig. \ref{fig5}a that the spirals are smaller
for larger rates $\kappa$, which is readily verified through the inspection of different snapshots.
The second length $L_T(t)$, which is a measure of the typical domain size, 
increases algebraically with time, as expected for a coarsening regime where larger domains grow at 
the expense of the smaller ones. For $\kappa = 0.7$ resp. $\kappa = 0.5$ the exponent governing the growth 
of this typical length takes on the value 0.47 resp. 0.43 \cite{Brown17}.
For $\kappa=0.3$ the algebraic regime is entered at a much later time, as seen in Fig. \ref{fig5}a, in accordance with the
fact that a smaller $\kappa$ yields larger spirals. 
As already noted in \cite{Brown17} the correlation length extracted from $C(t,r)$ for small values of $C_0$ also allows to 
monitor the growth of domains (see Fig. \ref{fig4} in \cite{Brown17}).
Different types of behavior are
also encountered in Fig. \ref{fig5}b for the densities of empty sites: whereas the density of empty sites formed in in-team reactions ($\rho(t)$, dashed lines) 
increases at a small rate as more and more spirals fit into the growing domains,
the density of empty sites formed at the boundaries between teams 
($\rho_T(t)$, full lines) decreases algebraically
with time (for $\kappa = 0.7$ and $\kappa =0.5$ one finds a value close to $-0.25$ for that exponent \cite{Brown17}) 
due to the shrinking total interface length during the
coarsening process. 

\begin{figure} [h]
\includegraphics[width=0.45\columnwidth,clip=true]{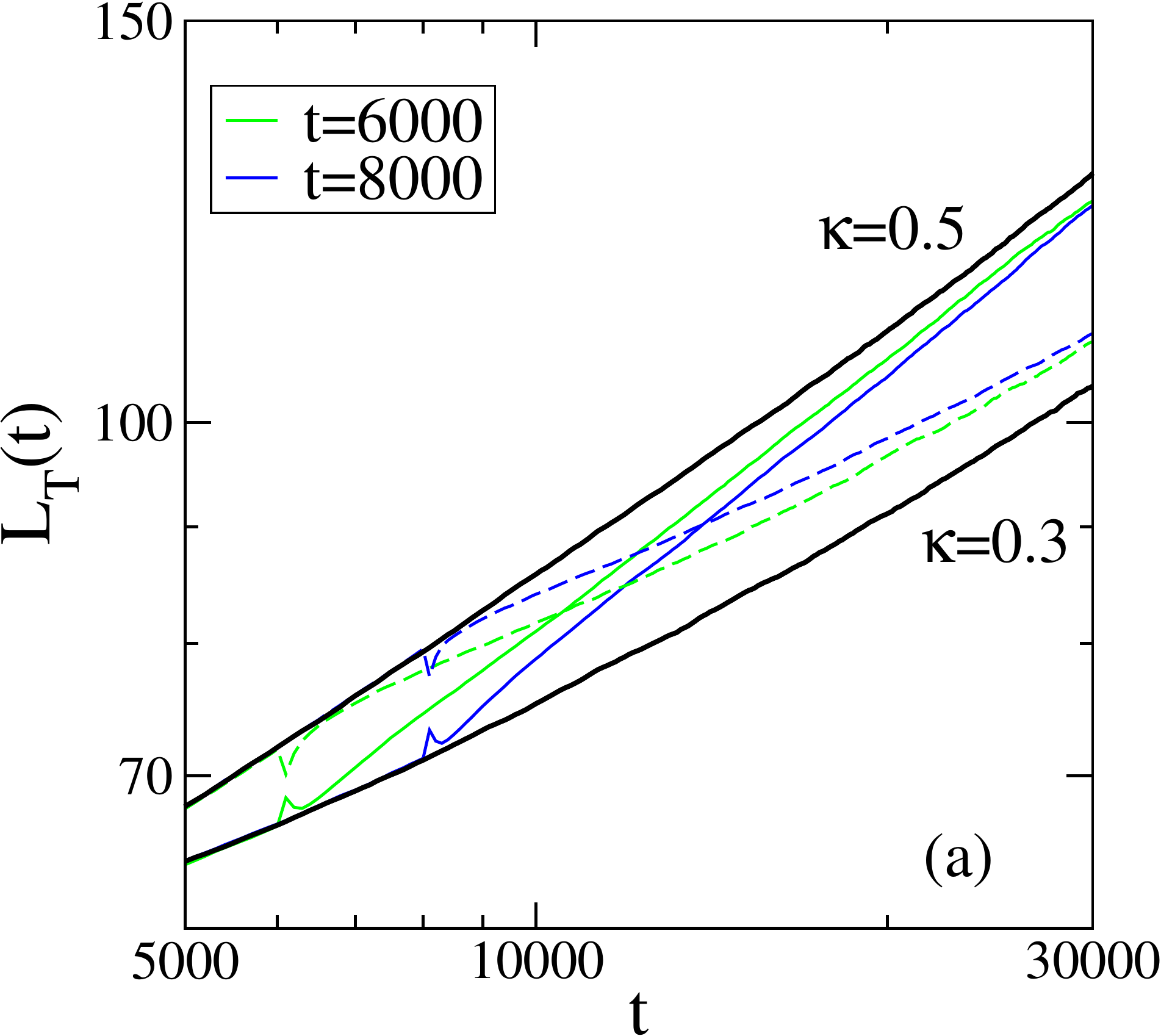}
\includegraphics[width=0.45\columnwidth,clip=true]{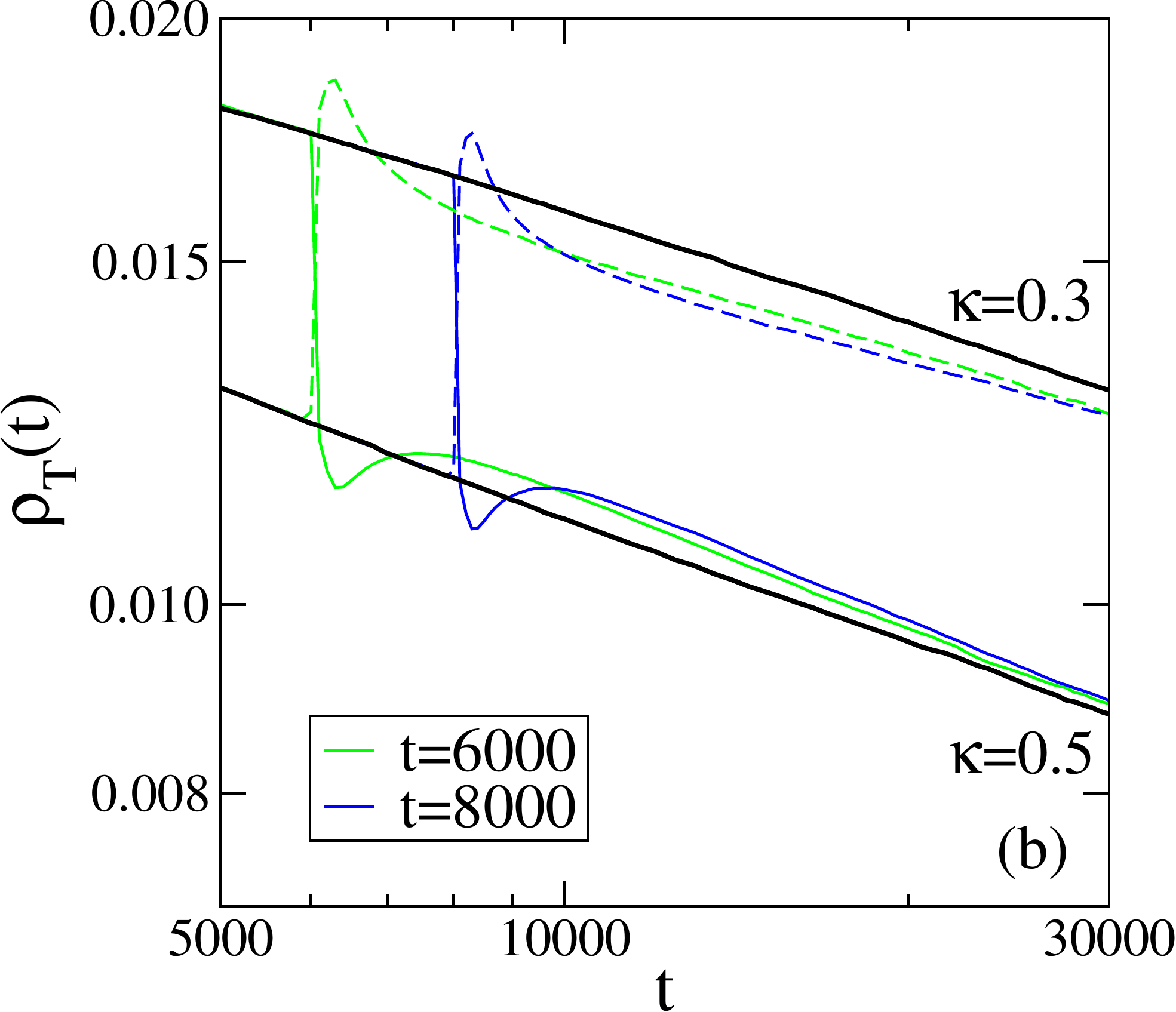}\\
\includegraphics[width=0.45\columnwidth,clip=true]{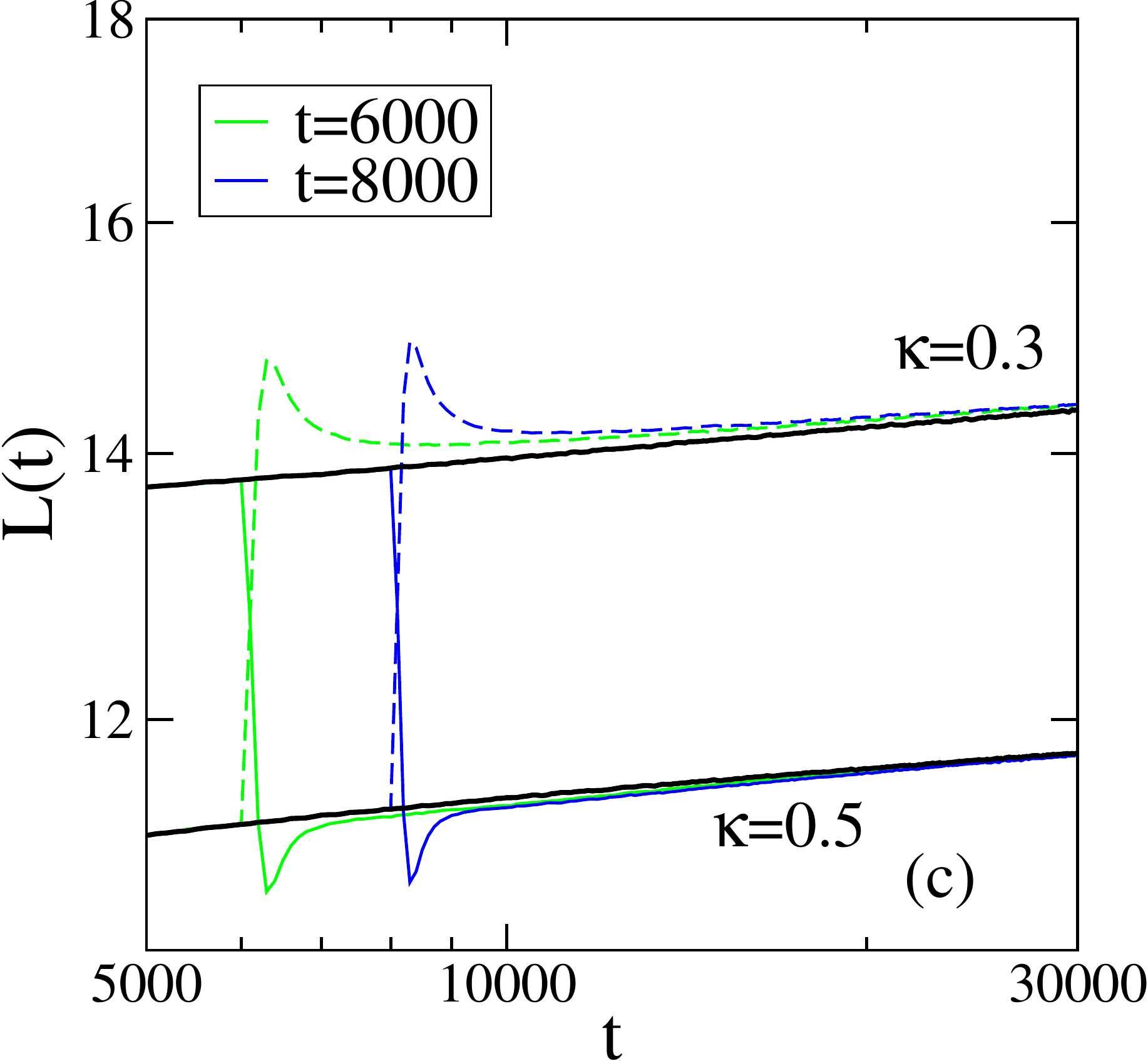}
\includegraphics[width=0.45\columnwidth,clip=true]{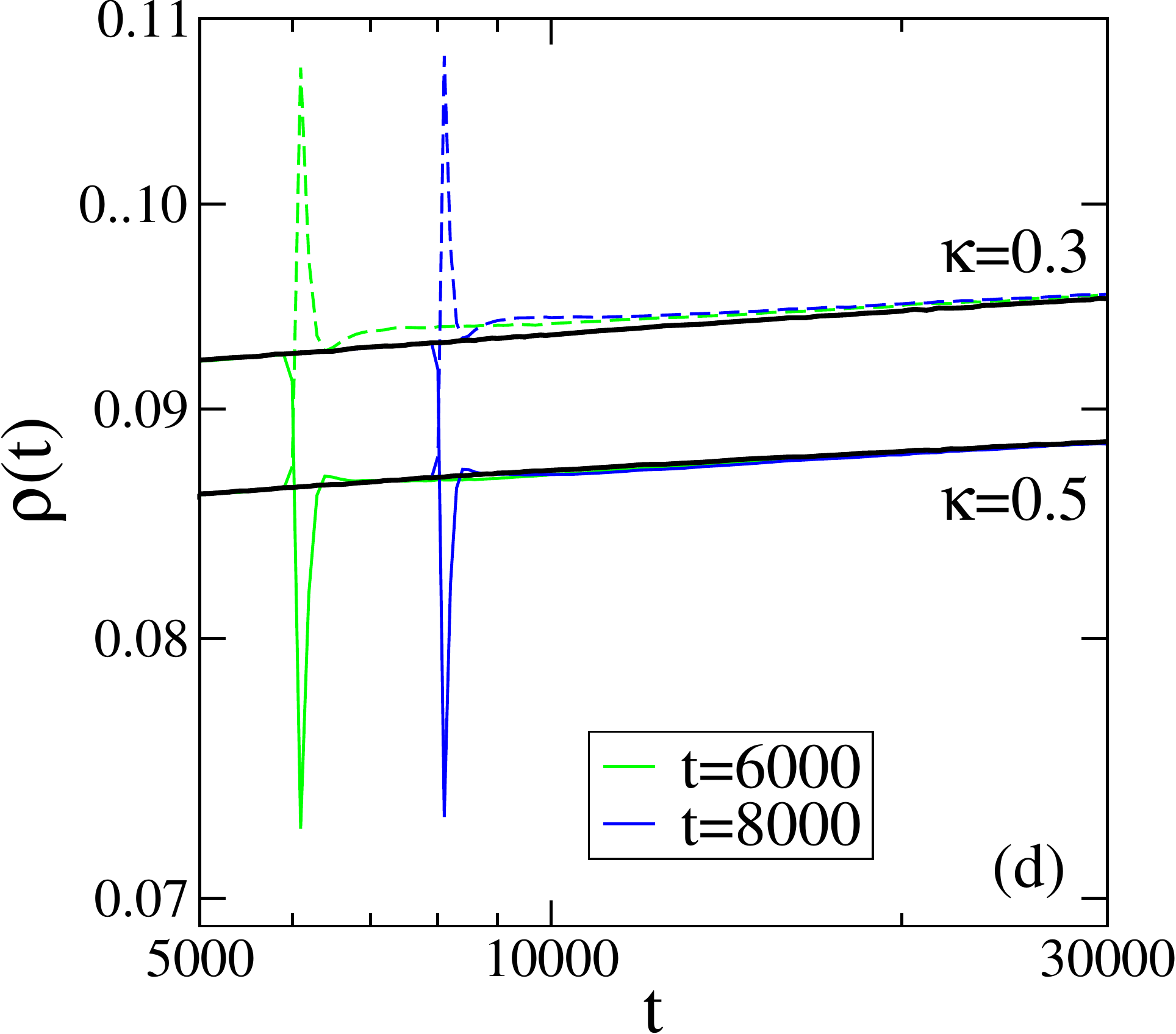}
\caption{\label{fig6} Time evolution of (a) the correlation length $L_T(t)$ obtained from the intersection of the
team-team space-time correlation function, (b)
the density of empty sites $\rho_T(t)$ created at the boundaries separating the different teams, (c)
the correlation length $L(t)$ extracted from the standard space-time correlation function, and (d)
the density of empty sites $\rho(t)$ that result from interactions between two members of the same team.
Thin full lines: changing the predation and reproduction rates from
$\kappa_i = 0.3$ to $\kappa_f = 0.5$, dashed lines: changing the predation and reproduction rates from
$\kappa_f = 0.5$ to $\kappa_i = 0.3$. The thick black lines are those of the unperturbed systems with fixed
rates $\kappa = 0.3$ resp. $\kappa = 0.5$. The simulated lattices have $700 \times 700$ sites, and the data
are the average of at least 4000 independent runs.
}
\end{figure}

We are now ready to discuss how the system adapts to a sudden change of the value of the rate from $\kappa = \kappa_i$ to
$\kappa = \kappa_f$. Recall that $\kappa$ is the value of both the predation rate and of the rate at which
offsprings are created on a neighboring empty site and that changing $\kappa$ also means changing the mobility
$\sigma = 1 - \kappa$. Fig. \ref{fig6} shows the results of two different protocols. In the first one, indicated
by the full green and blue lines, the rates are changed from $\kappa_i = 0.3$ to $\kappa_f =0.5$, whereas in the second one, see the dashed lines, the rates
are changed from $\kappa_i =0.5$ to $\kappa_f = 0.3$. The data shown in the figure have been obtained for changes taking place after $t=6000$ (green) or
$t=8000$ (blue) time steps. At these times the system has entered the regime where a typical domain is filled with multiple spirals. 

\begin{figure} [h]
\includegraphics[width=0.45\columnwidth,clip=true]{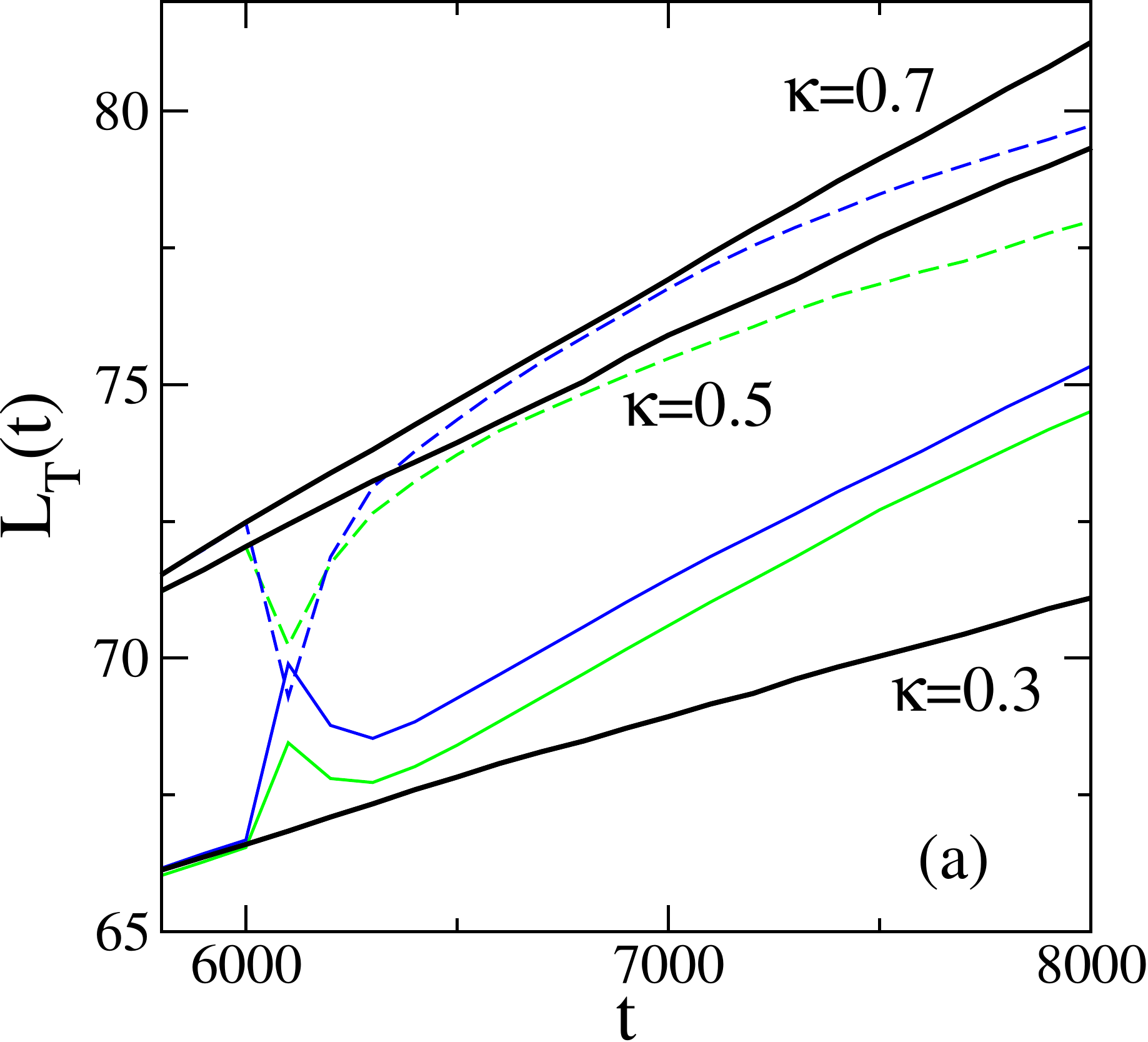}
\includegraphics[width=0.45\columnwidth,clip=true]{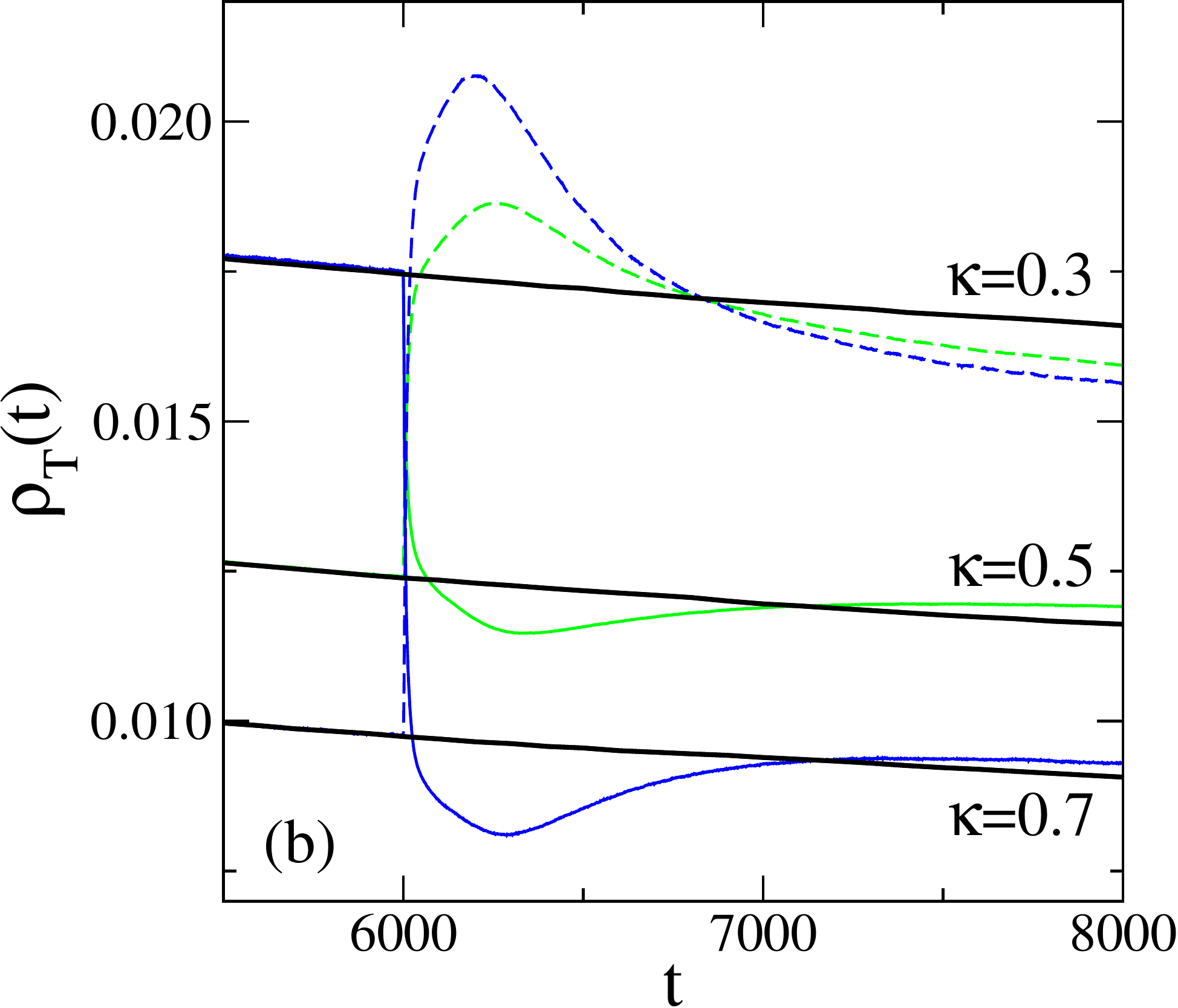}\\
\includegraphics[width=0.45\columnwidth,clip=true]{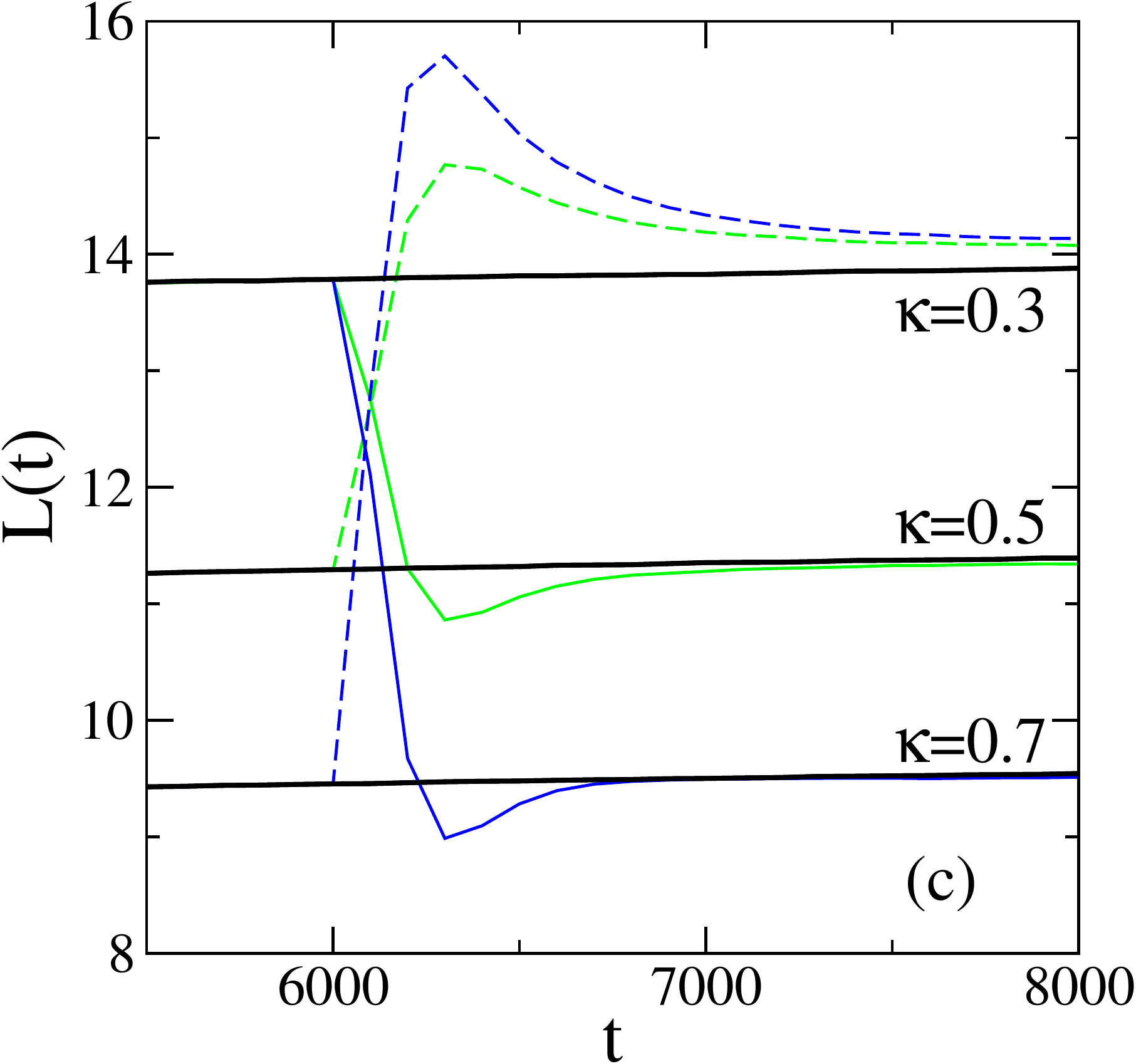}
\includegraphics[width=0.45\columnwidth,clip=true]{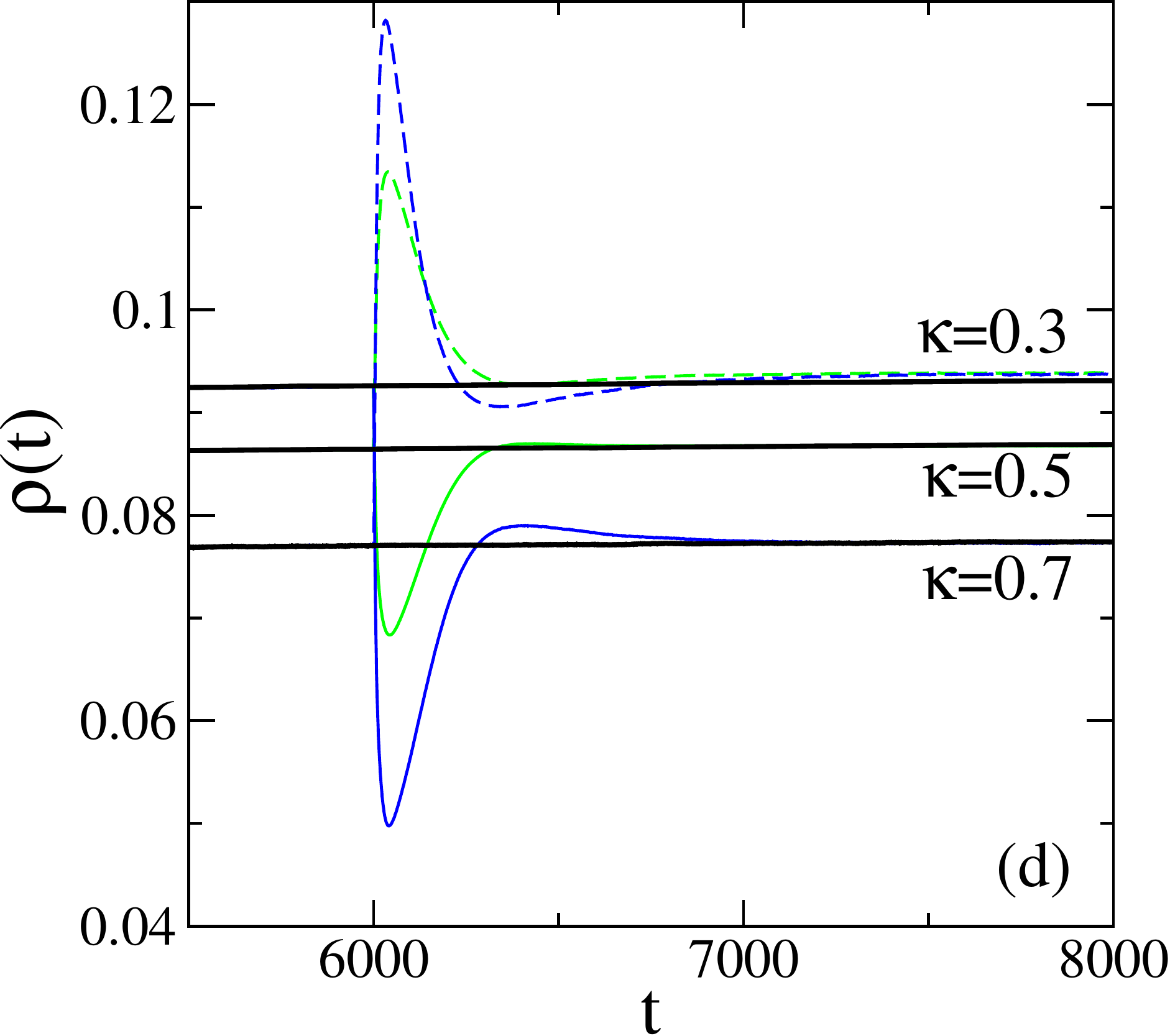}
\caption{\label{fig7} 
Comparison of the initial response for the four cases ($\kappa_i = 0.3$, $\kappa_f =0.5$) (full green (gray) lines), ($\kappa_i =0.5$, $\kappa_f = 0.3$)
(dashed green (dashed gray) lines), ($\kappa_i = 0.3$, $\kappa_f =0.7$) (full blue (thin black) lines), and ($\kappa_i =0.7$, $\kappa_f = 0.3$) (dashed blue (dashed black) lines)
when changing the rates at $t=6000$.
The different panels show the same quantities as in the corresponding panels in Fig. \ref{fig6}. 
The thick black lines are those of the unperturbed systems with fixed
rates $\kappa = 0.3$,  $\kappa = 0.5$, and $\kappa = 0.7$.
}
\end{figure}

The quick initial response observed in Fig \ref{fig6} is best understood when remembering that a smaller value of $\kappa$ results in a larger number
of empty sites, in smaller domains, and in larger spirals. If we then increase the value of $\kappa$, the system quickly fills
these additional empty sites, as witnessed by the sharp drop in the densities in Fig. \ref{fig6}b and \ref{fig6}d,
while at the same time the domain size increases and the spiral size decreases. 
When getting rid of this excess of empty sites, the system overshoots, and
the correction to this yields the observed non-monotonous behavior. The reverse effect is observed when decreasing $\kappa$: the smaller rate of reproduction
results in an immediate increase of empty sites and, concomitantly, in a quick decrease of the domain size and a rapid
increase of the spiral size. The system also overshoots in this
case, as shown in Fig. \ref{fig6}.

In Fig. \ref{fig7} we compare the initial response for four different cases: 
($\kappa_i = 0.3$, $\kappa_f =0.5$) (full green lines), ($\kappa_i =0.5$, $\kappa_f = 0.3$)
(dashed green lines), ($\kappa_i = 0.3$, $\kappa_f =0.7$) (full magenta lines), and ($\kappa_i =0.7$, $\kappa_f = 0.3$) (dashed magenta lines).
Inspection of the different panels reveals that the initial response is more violent the larger the value of  $\left| \kappa_i - \kappa_f \right|$
is. The two cases ($\kappa_i =0.5$, $\kappa_f = 0.3$) and ($\kappa_i =0.7$, $\kappa_f = 0.3$) having the same final value of the rate $\kappa$,
we can easily compare for these cases (dashed lines in Fig. \ref{fig7}) the magnitudes by which the systems overshoot. For 
the empty site densities as well as for $L(t)$
the difference between the height of the maximum and the line from the unperturbed system at $\kappa = 0.3$ increases by a factor close to 2
when replacing $\kappa_i = 0.5$ by $\kappa_i=0.7$. 
Keeping $\kappa_f$ at 0.3 and increasing $\kappa_i$ from 0.4 to 0.9, we find that the
overshoots for the different quantities all first increase algebraically with $\left| \kappa_i - \kappa_f \right|$ with exponents
close to 1, before saturating when $\kappa_i$ approaches 1. These limit values (which is 2.15 for $L(t)$ and $\kappa_f=0.3$) 
depend on the value of $\kappa_f$.

After this quick initial response 
to the perturbation, the properties of the system evolve to those of the unperturbed system with constant $\kappa = \kappa_f$.
Whereas the quantities related to the in-domain dynamics (correlation length $L(t)$ and density of empty sites $\rho(t)$
due to in-team reactions) adjust rather quickly to the change, as shown in Fig. \ref{fig6}c and \ref{fig6}d, the quantities related
to the coarsening domains, see Fig. \ref{fig6}a and \ref{fig6}b, take more time to approach the lines of the unperturbed system.
Calling $L_{T,\kappa_f}$ resp. $\rho_{T,\kappa_f}(t)$ the corresponding quantities from the unperturbed systems,
the distances $\left| L_T(t) - L_{T,\kappa_f}(t) \right|$ and $\left| \rho_T(t)- \rho_{T,\kappa_f}(t)\right|$ approach zero algebraically, with exponents 
$x_L$ and  $x_\rho$, i.e. $\left| L_T(t) - L_{T,\kappa_f}(t) \right| \sim t^{-x_L}$ and $\left| \rho_T(t)- \rho_{T,\kappa_f}(t)\right| 
\sim t^{-x_\rho}$. For the cases shown in Figs. \ref{fig6} and \ref{fig7} we have 
$x_L=0.68(5)$ and $x_\rho =  1.45(7)$ for ($\kappa_i=0.3$, $\kappa_f =0.5$), $x_L=0.37(2)$ and $x_\rho =  1.06(5)$ for ($\kappa_i=0.3$, $\kappa_f =0.7$),
$x_L=0.39(3)$ and $x_\rho = 0.88(3)$ for ($\kappa_i=0.5$, $\kappa_f =0.3$), and
$x_L=0.30(3)$ and $x_\rho = 0.96(3)$ for ($\kappa_i=0.7$, $\kappa_f =0.3$). In general, we observe that the  
exponents are larger for $\kappa_i < \kappa_f$ than for $\kappa_i > \kappa_f$.
In the former case, domains are undersized once $\kappa$ has been increased, and the system adjusts to this 
by going through an
accelerated growth phase. When $\kappa_i > \kappa_f$, the domains are too large after the decrease of $\kappa$. This, however, does not interrupt
the growth process, but only slows down the domain growth. This behavior is similar to that observed previously in the two-dimensional $ABC$ model
when changing the value of the swapping rate during the coarsening process \cite{Brown15}.

While we focused here on the values 0.3, 0.5, and 0.7 for $\kappa$, we checked that the same behavior is encountered for other values of $\kappa_i$
and $\kappa_f$. We also note that we did not investigate the reaction of the system to rate changes that take place at much earlier times, i.e. at times before
the coarsening of domains with multiple spirals sets in. As it is difficult to interpret the response of the system in a regime where the space-time patterns
have not yet fully formed, we refrain from discussing this here.

\section{Changing the scheme}

A perturbation like the change of reaction rates discussed in the previous Section does not disrupt space-time patterns.
A more brutal perturbation is the change of the reaction scheme as it forces the different
species into new relationships, with new alliances replacing earlier ones. In a spatial setting, changes to the
partnerships yield changes to the emerging space-time structures.

We discuss in the following how six species playing a (6,3) game adjust when the scheme is suddenly changed to (6,2).
This is considered to be a transient perturbation that only lasts for a specified number of time steps before 
the scheme reverses back to the (6,3) interactions.
If the perturbation lasts long enough, the system has time to fully adapt to the new situation before the second change
of scheme takes place.

\subsection{The well-mixed case}

\begin{figure} [h]
\includegraphics[width=0.45\columnwidth,clip=true]{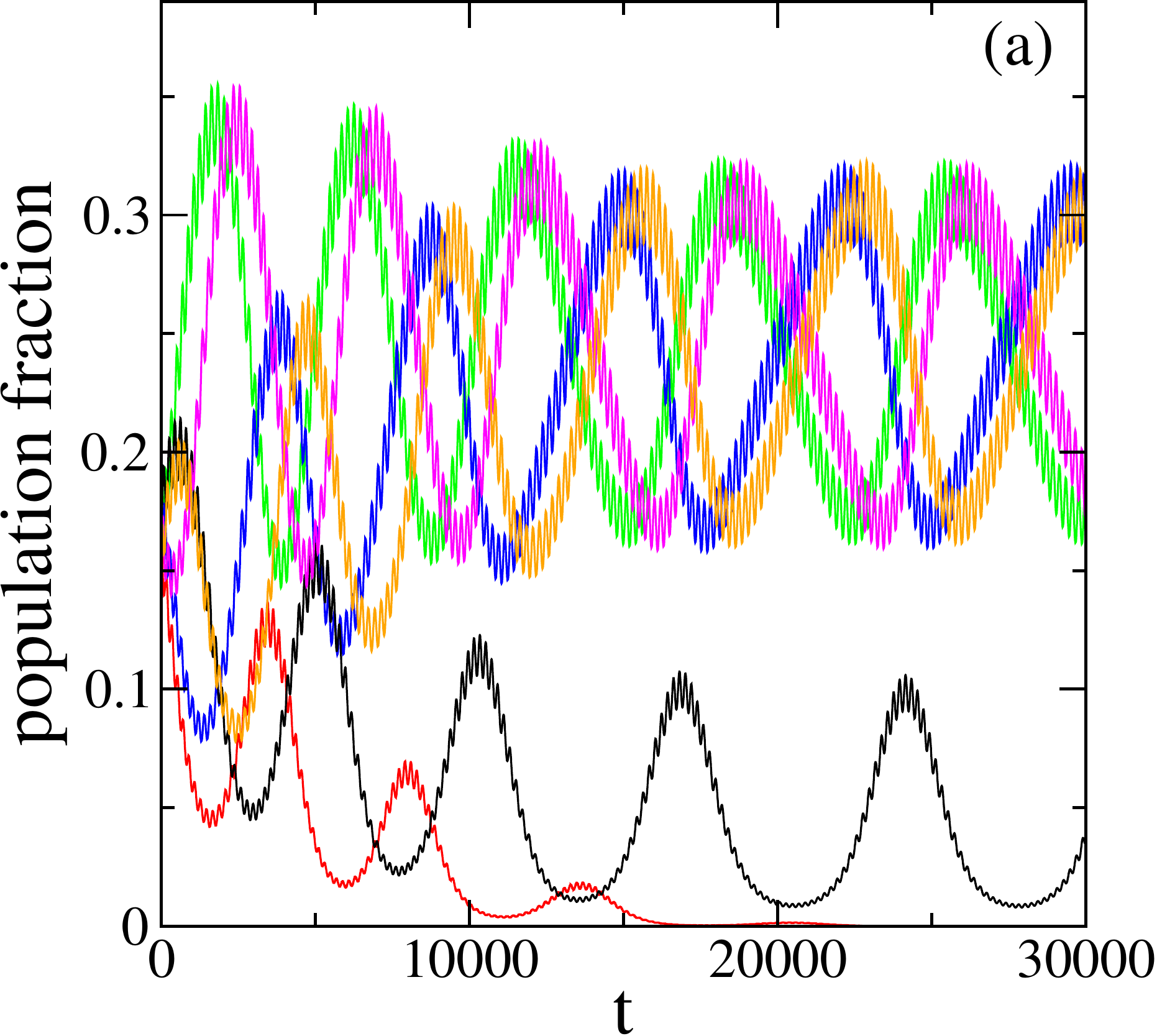}
\includegraphics[width=0.45\columnwidth,clip=true]{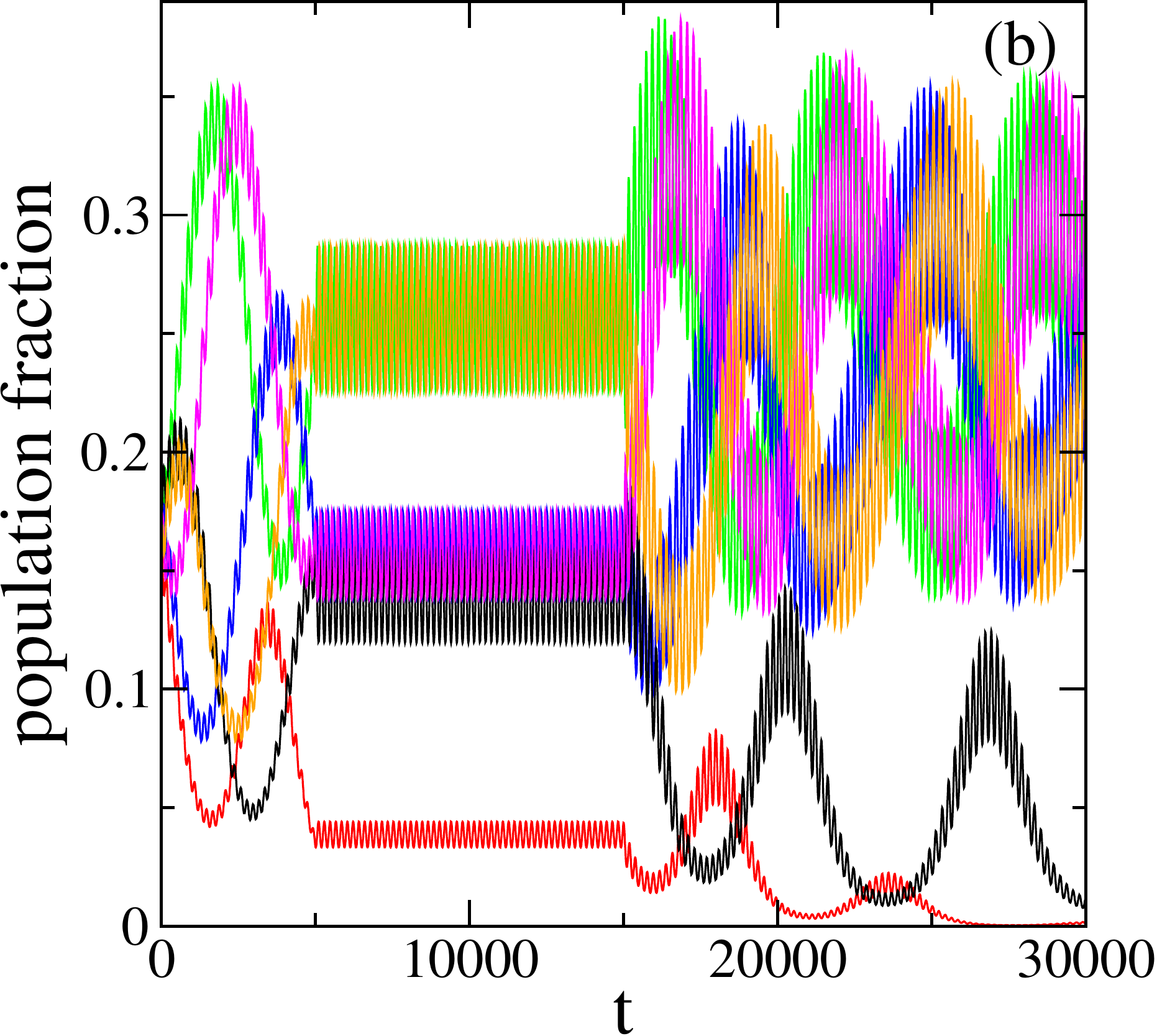}\\
\includegraphics[width=0.45\columnwidth,clip=true]{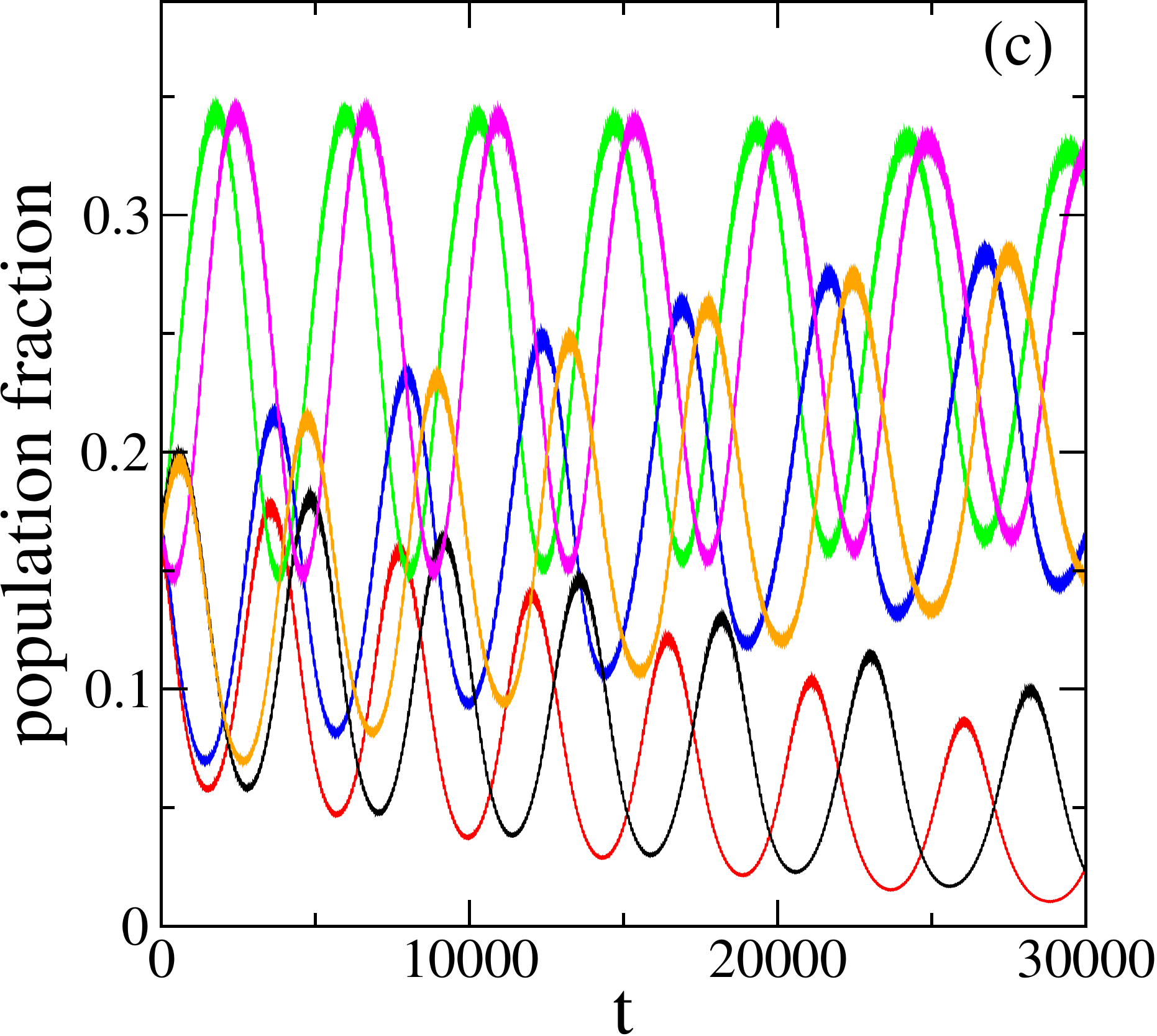}
\includegraphics[width=0.45\columnwidth,clip=true]{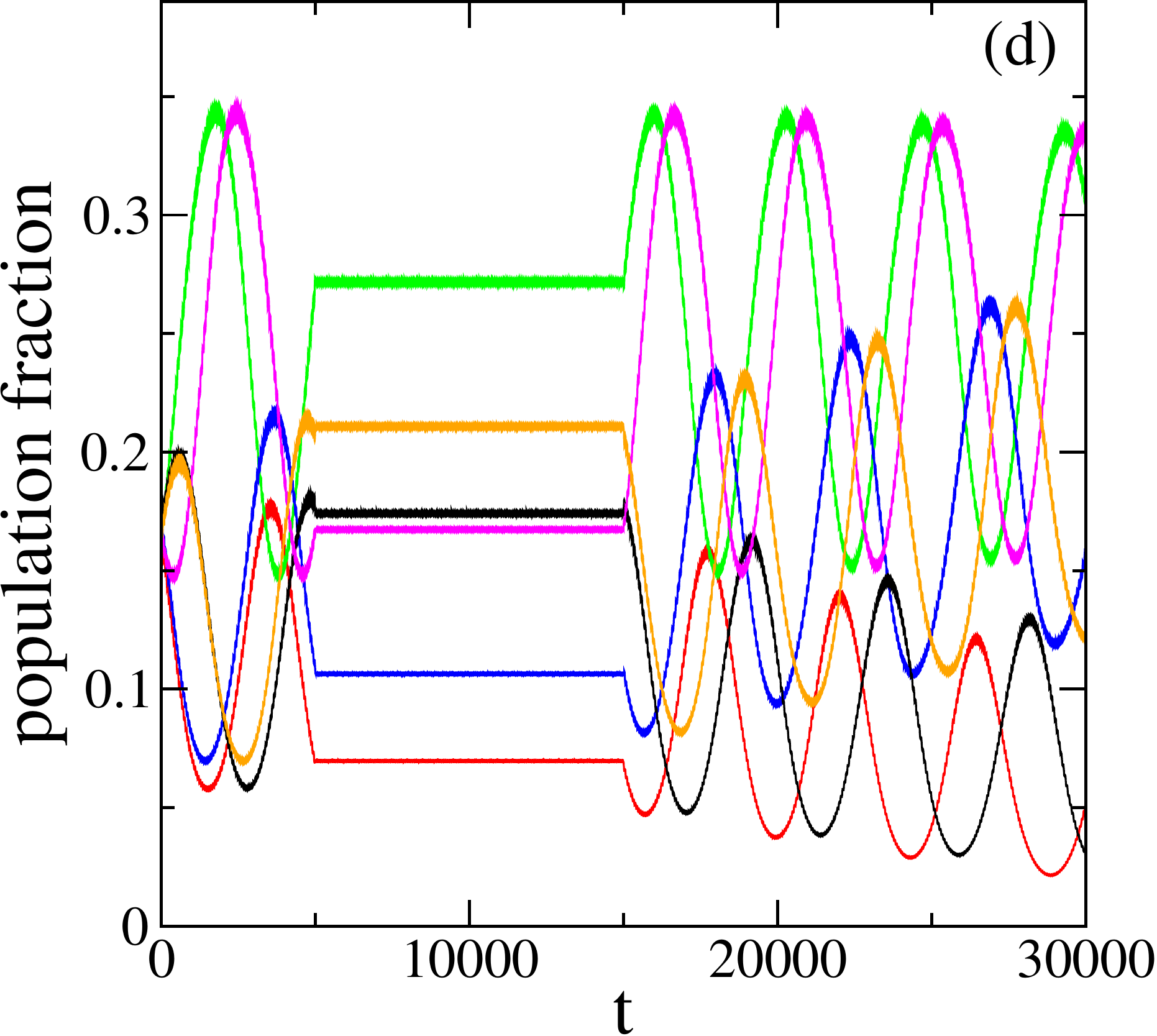}
\caption{\label{fig8} 
(a) Time dependence of the population fractions as obtained from the rate equations (\ref{rateeq}) with the values 0.06 for all rates,
with the exception of the rates $\kappa_{41}$, $\kappa_{52}$, and $\kappa_{63}$ that have been set to 0.07. (b) Time evolution
of the population fractions when at time $t=5000$ the scheme is changed from (6,3) to (6,2). The rates are the same as for (a).
This perturbation is kept for 10000 time steps. (c) and (d) are the same as (a) and (b), but now for the rates 0.2 and 0.21.
In both cases the system has been set up with equal number
densities 1/6 for all species. The fourth order Runge-Kutta integration scheme with step length $\Delta t = 0.01$ 
has been used. The two species for which the population fractions approach zero are species 1 in red (dark gray) and species 4 in black.
}
\end{figure}

For the well-mixed case without spatial setting we consider the mean-field level description of rate equations.
Calling $n_1$ the population fraction of species 1 (and similar for the other species), the equation describing the time
evolution of $n_1$ is
\begin{equation} \label{rateeq}
\frac{d n_1}{d t} = \kappa_{12} n_1 n_2 + \kappa_{13} n_1 n_3 + \left( \kappa_{14} - \kappa_{41} \right) n_1 n_4 - \kappa_{51}
n_1 n_5 - \kappa_{61} n_1 n_6~,
\end{equation}
where $\kappa_{ij}$ is the rate at which the predator $i$ attacks the prey $j$. We immediately notice that the term
proportional to $n_1 n_4$ vanishes for the case $\kappa_{14} = \kappa_{41}$, resulting in the rate equations for the $(6,2)$
game. In order to avoid this we choose rates such that $\kappa_{14} \ne \kappa_{41}$. This of course means that the symmetry
between species is lost, as some species have an advantage over others. The data discussed in Fig. \ref{fig8}a 
and Fig. \ref{fig8}b have been obtained for rates 0.06,
with the exception of $\kappa_{41} = \kappa_{52}= \kappa_{63} = 0.07$. As shown in Fig. \ref{fig8}c and 
Fig. \ref{fig8}d for the rates 0.2 and 0.21, these results do not change qualitatively when 
choosing other values of the rates while keeping the advantage that species 4 has over species 1 etc.

As shown in Fig. \ref{fig8}a, the (6,3) scheme results in slow oscillations with large amplitudes that are accompanied by much faster
oscillations with small amplitudes. The slow oscillations with large amplitudes result from the rock-paper-scissors interactions
between the three species forming one alliance, see Fig. \ref{fig1}, and are specific to the (6,3) scheme. The period of the quick oscillations
are very similar to those encountered in the (6,2) game and manifest themselves due to the near cancellations of the interactions between
species $i$ and species $i+3$ (mod 6). Finally, due to the bias in the rates, species 1 is at a major disadvantage which results 
for this species in the decrease over time of the amplitude of the slow oscillations (red curve). Obviously, for the purpose of
this study we would prefer not having this bias, but this is the only way to investigate the transition between the (6,3) and (6,2)
schemes at the level of the rate equations.

\begin{figure} [h]
\includegraphics[width=0.45\columnwidth,clip=true]{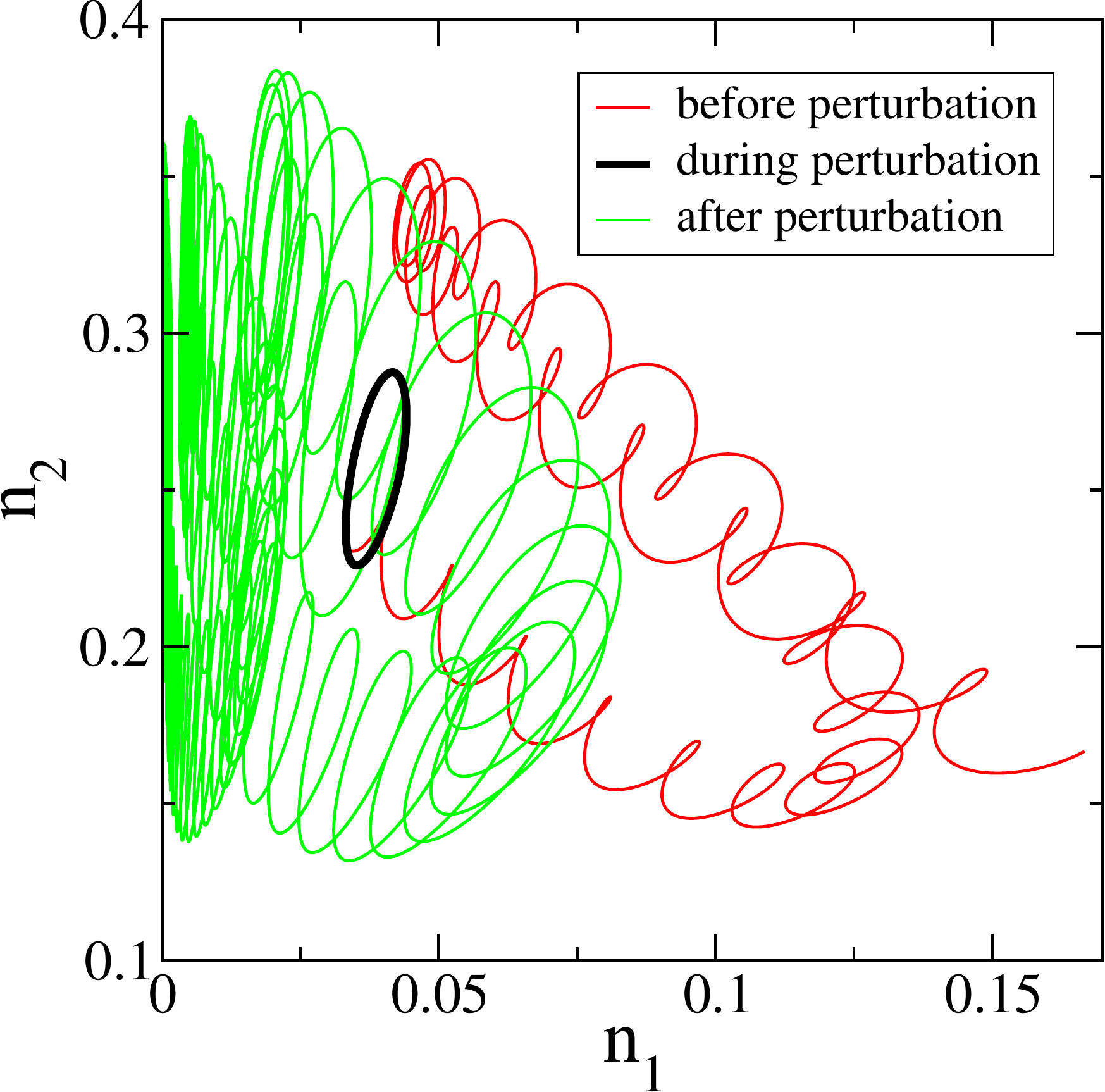}
\caption{\label{fig9} Trajectory in the $n_1-n_2$ phase space when changing between $t=5000$ and $t=15000$ the scheme
from (6,3) to (6,2), see Fig. \ref{fig8}b. Red (thin black) line: before the perturbation, thick black line: during the perturbation,
green (gray) line: after the perturbation.
}
\end{figure}

In Fig. \ref{fig8}b we show the time evolution of the population fractions where at time $t=5000$ we switch the scheme to $(6,2)$. As a result
of that switch the density of each species oscillates around the value it had at the moment of the switch. At a later time ($t=15000$ in the example of the 
figure), we change the scheme back to (6,3). The system quickly reestablishes the slow oscillations, but with an initial period that differs
slightly from that of the unperturbed (6,3) system shown in Fig. \ref{fig8}a. At the same time the amplitudes of the quick oscillations
are enhanced. Fig. \ref{fig9} provides a different view of how the system adapts to the modification of the interaction scheme 
through the trajectory in the $n_1-n_2$ phase space. The increase of the amplitude of the quick oscillations is readily seen in this plot too.

In Fig. \ref{fig8}c and \ref{fig8}d we repeat this study, but now for rates 0.2, with the exception of $\kappa_{41} = \kappa_{52}= \kappa_{63} = 0.21$.
Whereas no qualitative changes are observed, there are quantitative differences that show up when changing the rates. For example, the periods
of both the slow and fast oscillations decrease when increasing the predation rates. At the same time the amplitude of the fast oscillations
decrease when increasing $\kappa$. This is especially true for the time interval during which the interaction scheme (6,3) is replaced
by the scheme (6,2).

We also studied this situation through numerical simulations. Systems containing hundreds of thousands of individuals very closely follow
the trajectories obtained from the rate equations, with rapid adjustments to the new time pattern when the scheme is changed. As we will see
in the following, the same protocol applied to the spatial system results in complex transitions between the very distinctive
space-time patterns that characterize the (6,3) and (6,2) schemes.

\subsection{The spatial situation}

In two space dimensions both the (6,3) and the (6,2) reaction schemes result in the formation of alliances and domain coarsening, but
the composition of the alliances and the dynamics that takes place within the domains are very different, see Figs. \ref{fig3} and \ref{fig4}.
Whereas the (6,3) scheme yields two alliances with the three teams in each alliance undergoing a rock-paper-scissors game, the (6,2) game
results in the formation of three teams of two neutral partners.

We discuss in the following two different protocols. First we consider the case that an initially disordered system evolves for some time
following the (6,3) rules before we suddenly change the interaction scheme to (6,2). This scheme is kept until at a later time we change back to 
(6,3). Changing from (6,3) to (6,2) and back to (6,3) allows us to investigate the spatial and temporal signature for two different
changes of the interaction scheme. In the second protocol we periodically change between the (6,3) and (6,2) schemes and study how this affects
the ordering process. This protocol mimics the periodic changes in the interactions of species due to seasonal variations.

\begin{figure} [h]
\minipage{0.22\textwidth}
  \centering
  \includegraphics[width=\linewidth]{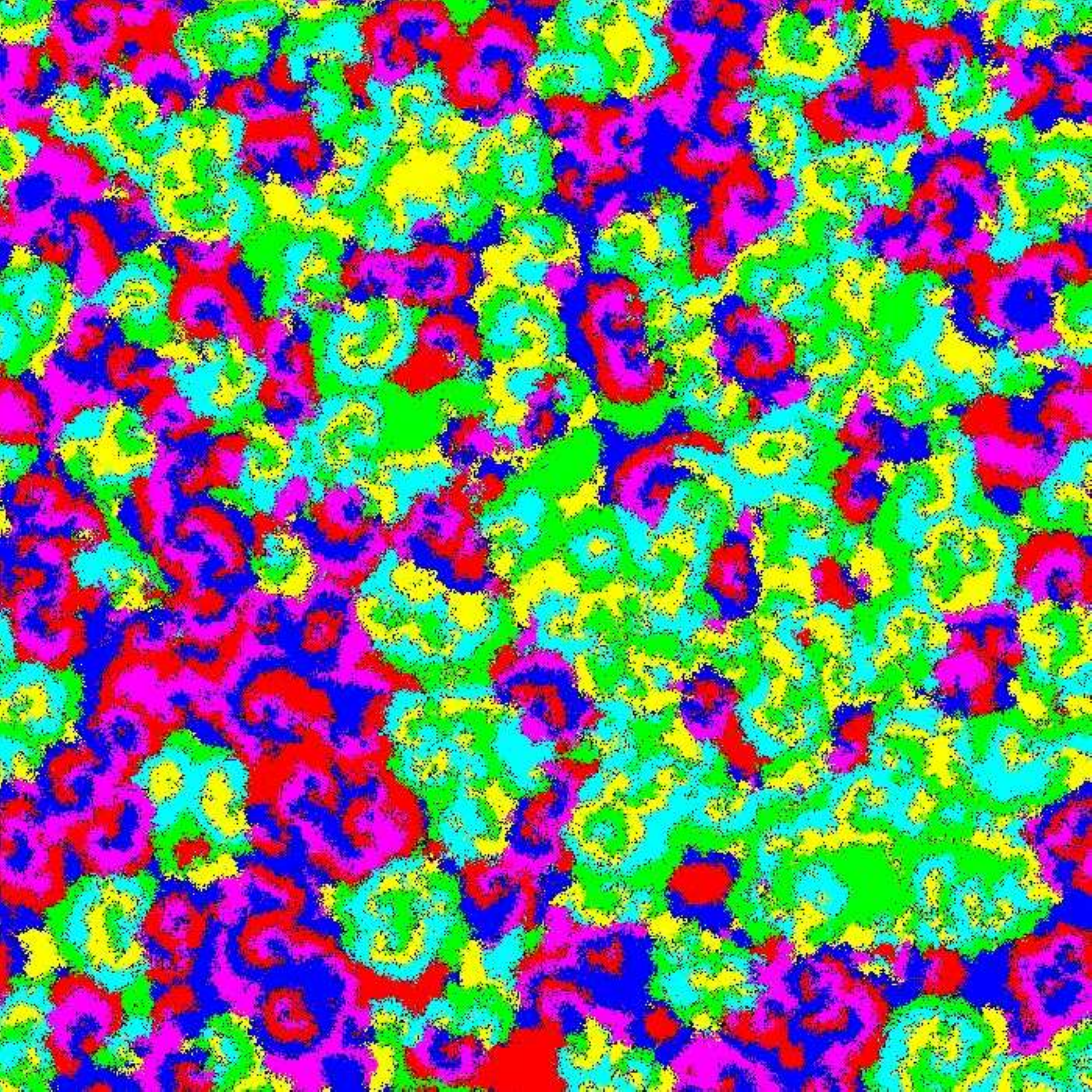}\\
  t=500
\endminipage
\hspace*{0.03\textwidth}
\minipage{0.22\textwidth}
  \centering
  \includegraphics[width=\linewidth]{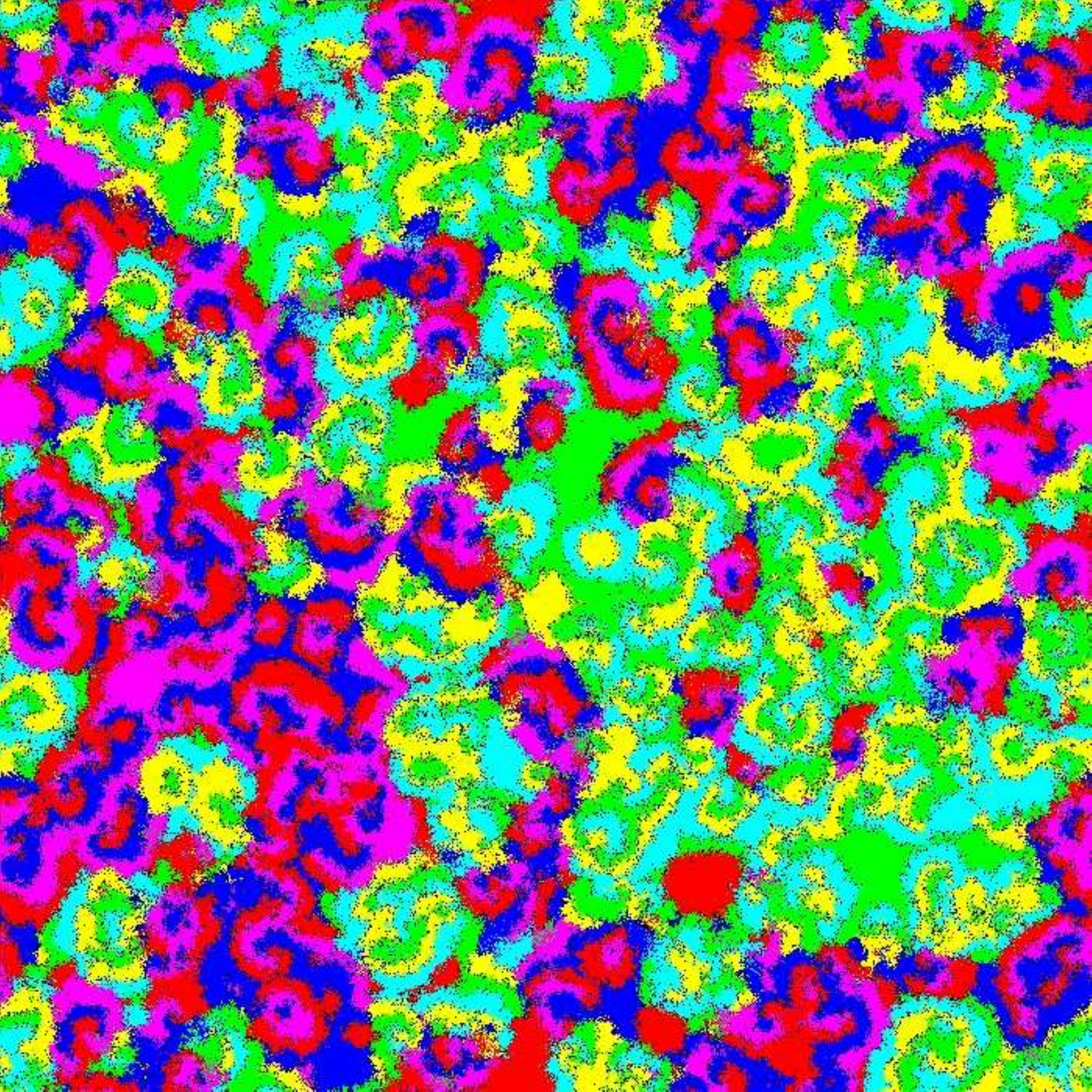}\\
  t=520
\endminipage
\hspace*{0.03\textwidth}
\minipage{0.22\textwidth}
  \centering
  \includegraphics[width=\linewidth]{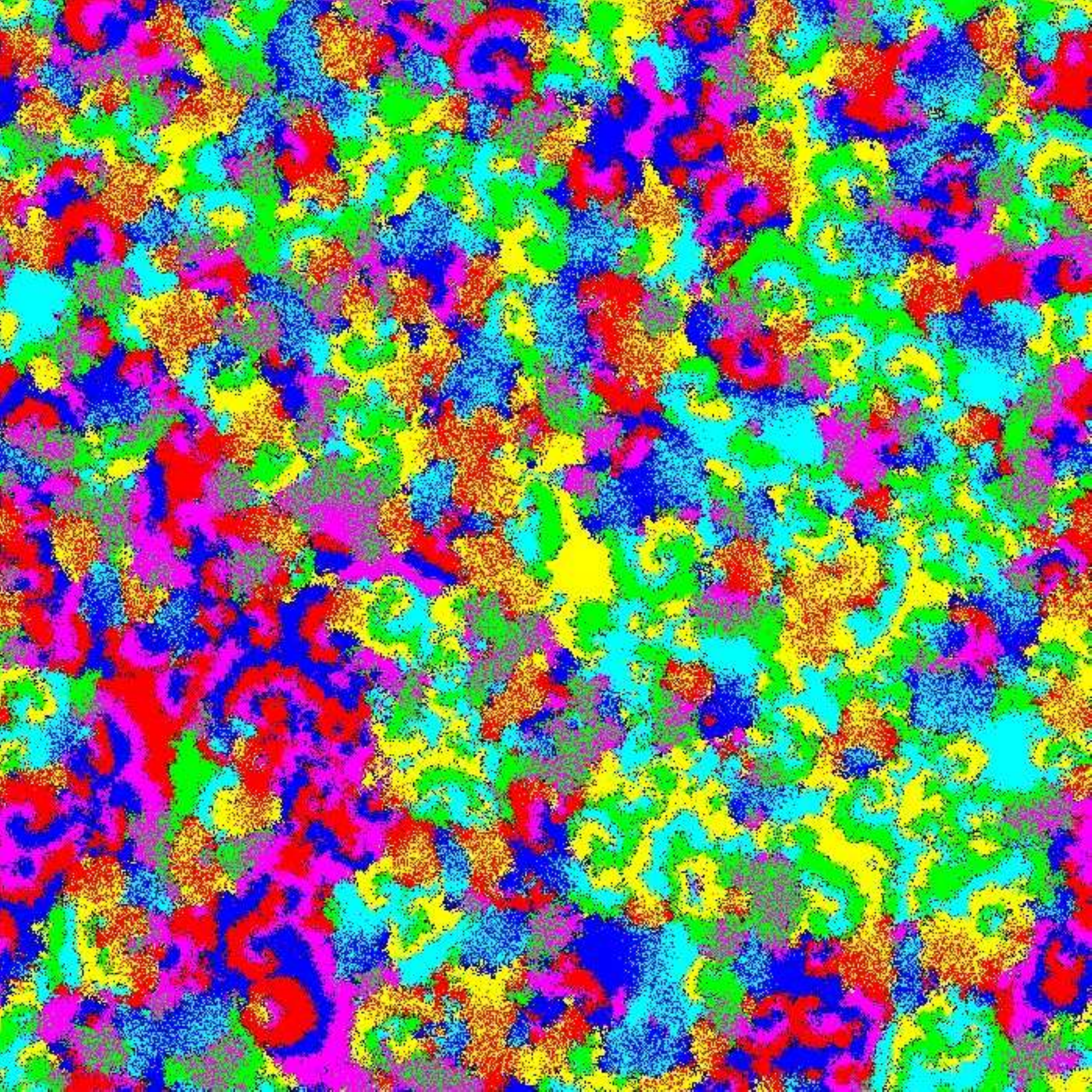}\\
  t=600
\endminipage
\hspace*{0.03\textwidth}
\minipage{0.22\textwidth}
  \centering
  \includegraphics[width=\linewidth]{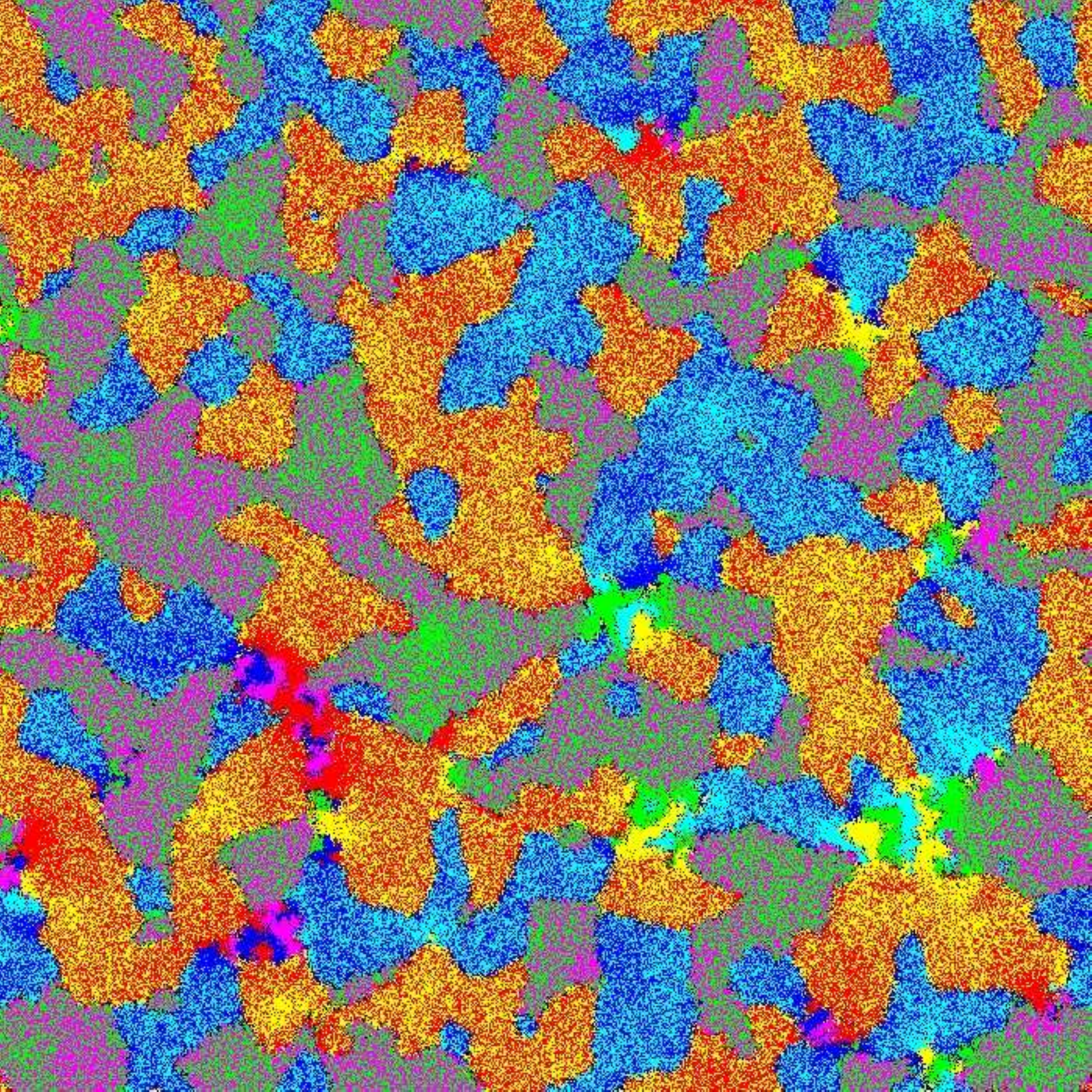}\\
  t=750
\endminipage\\[0.5cm]
\minipage{0.22\textwidth}
  \centering
  \includegraphics[width=\linewidth]{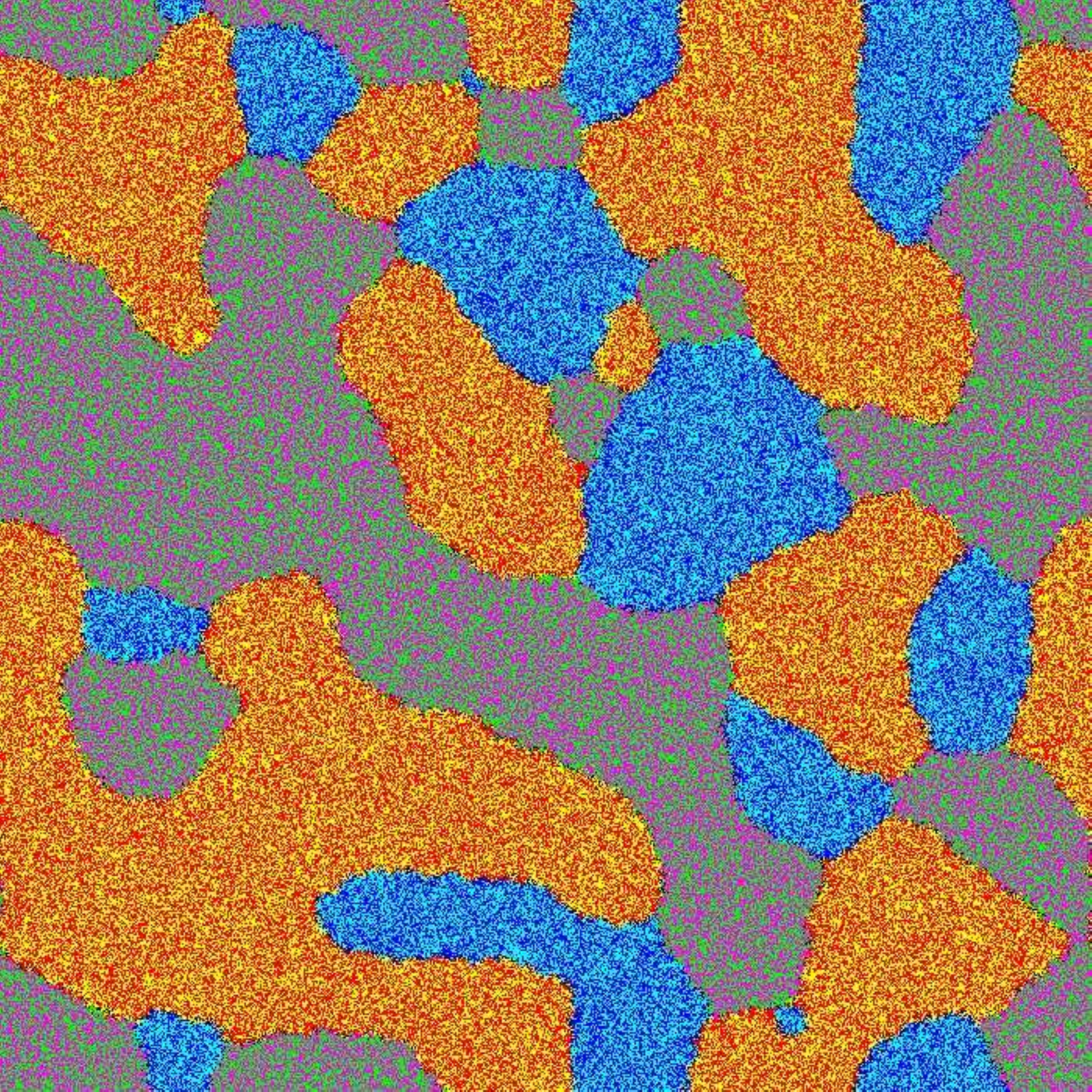}\\
  t=4500
\endminipage
\hspace*{0.03\textwidth}
\minipage{0.22\textwidth}
  \centering
  \includegraphics[width=\linewidth]{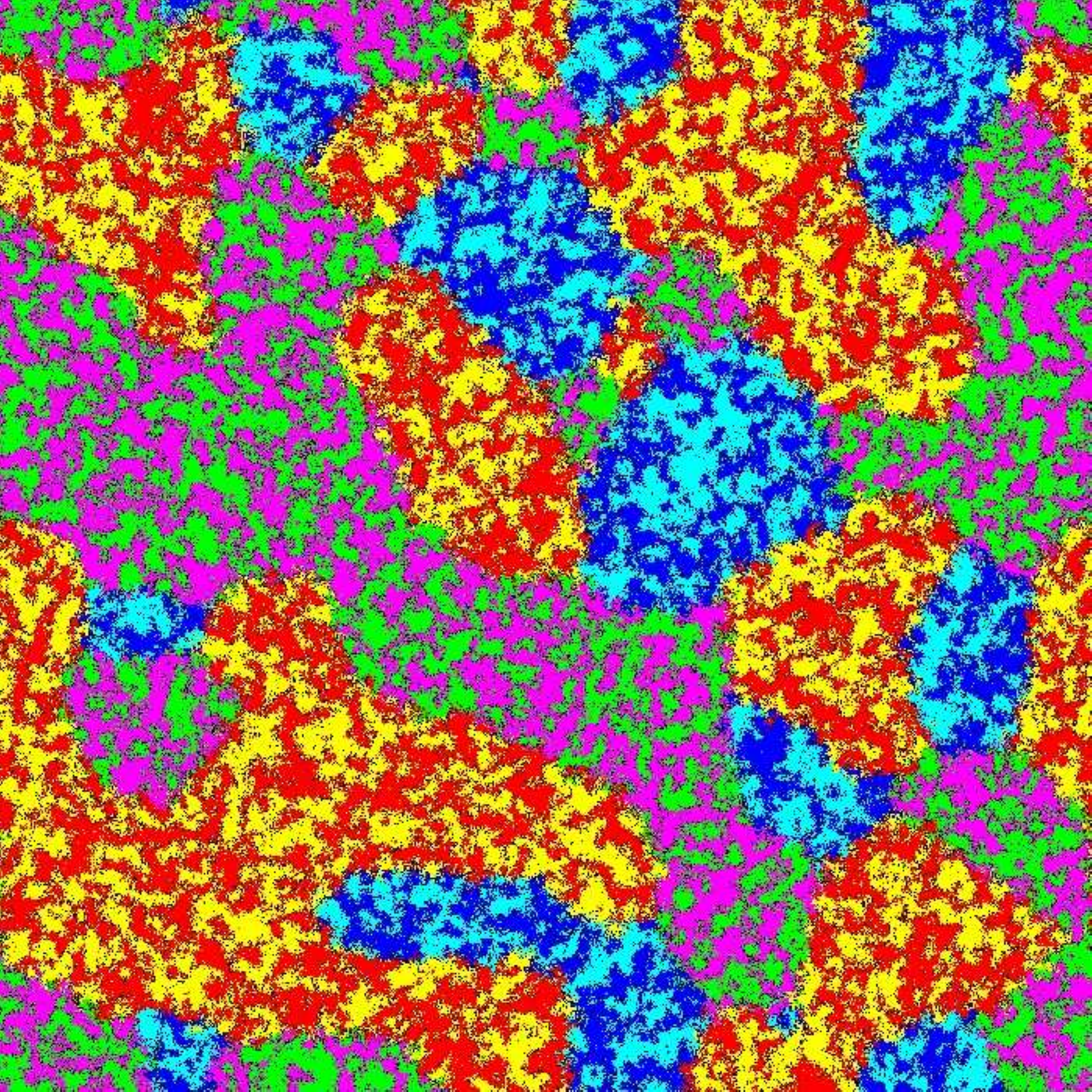}\\
  t=4520
\endminipage
\hspace*{0.03\textwidth}
\minipage{0.22\textwidth}
  \centering
  \includegraphics[width=\linewidth]{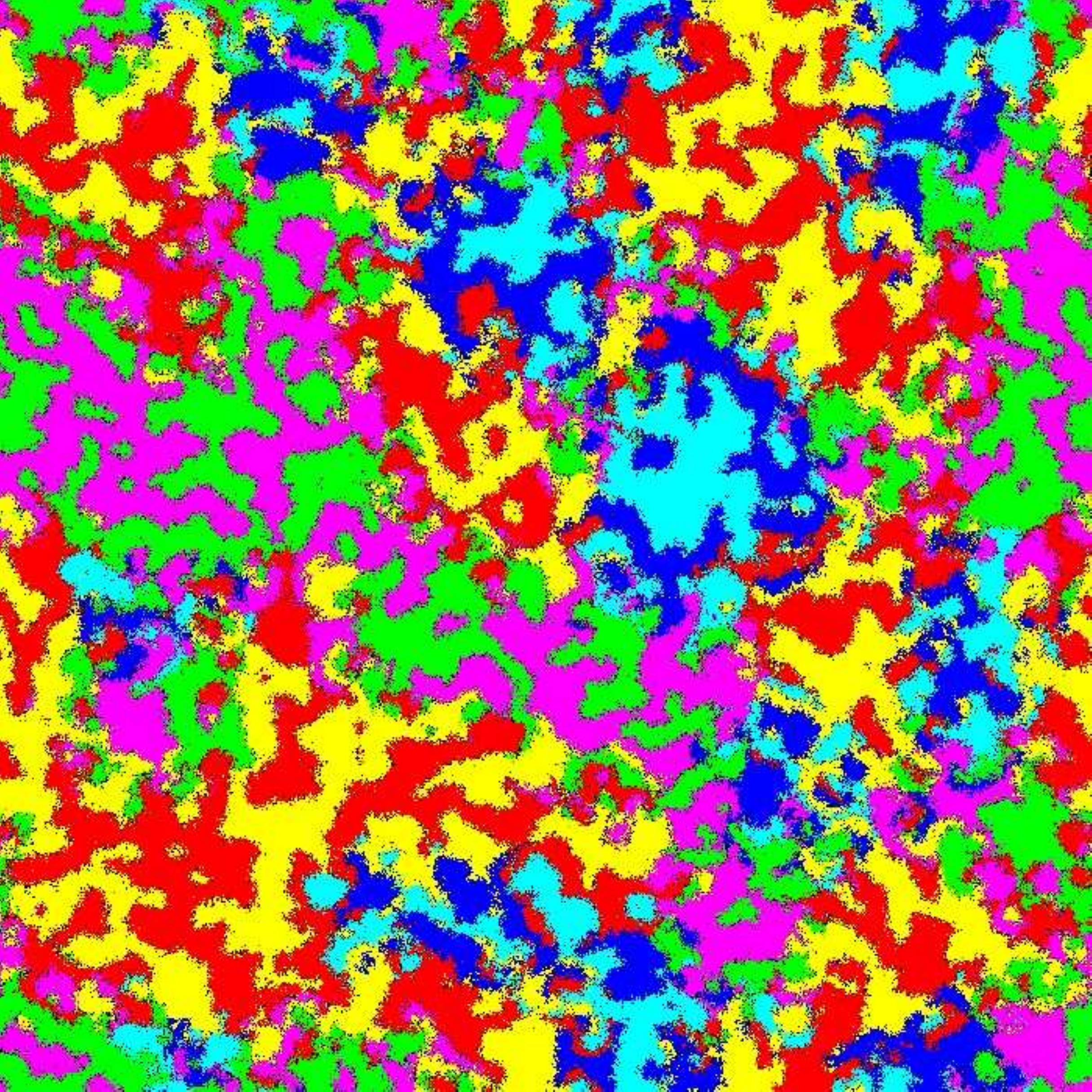}\\
  t=4600
\endminipage
\hspace*{0.03\textwidth}
\minipage{0.22\textwidth}
  \centering
  \includegraphics[width=\linewidth]{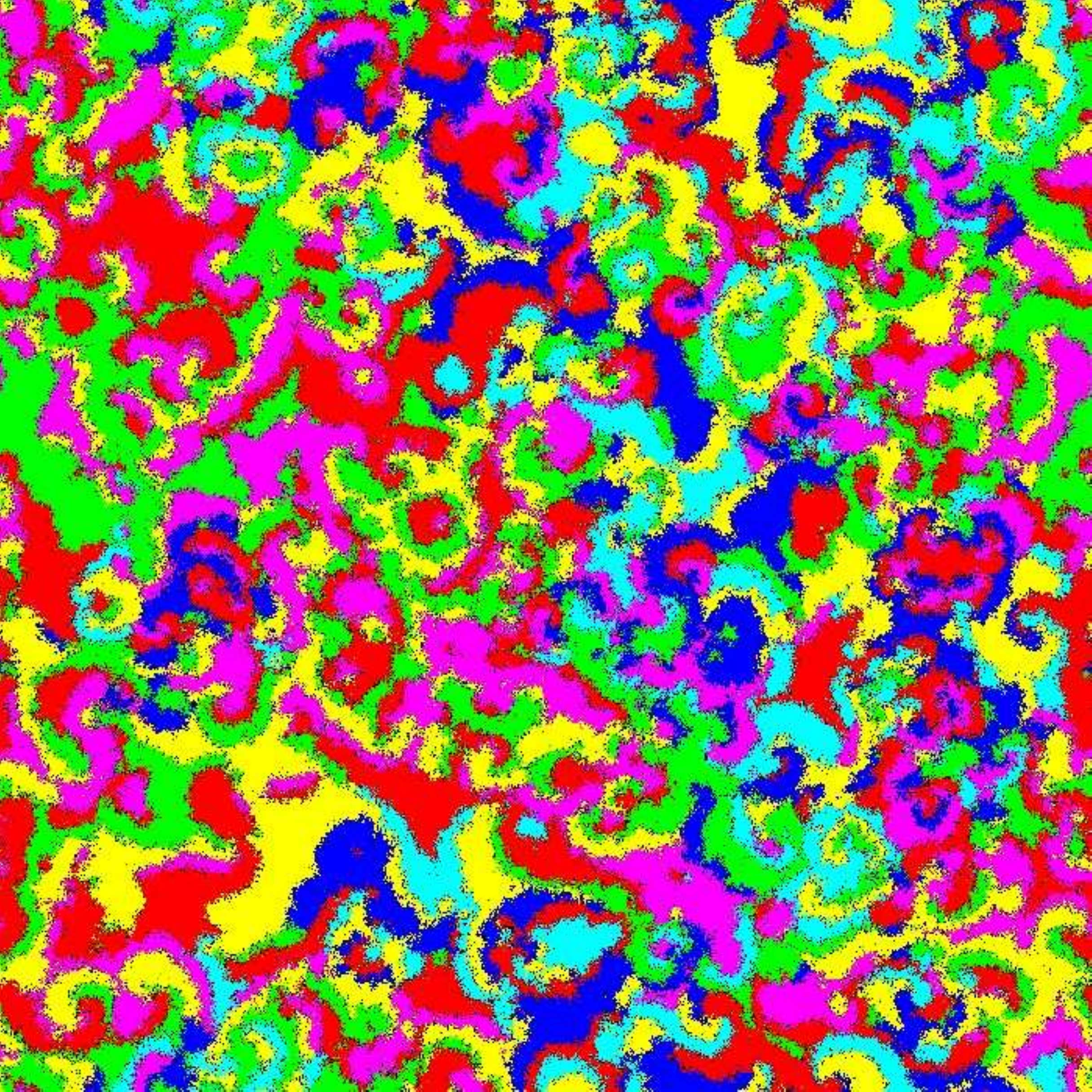}\\
  t=4750
\endminipage\\
\caption{\label{fig10} 
Snapshots of the system composed of $700 \times 700$ sites with $\kappa = 0.5$ where at $t=500$ the interaction scheme is changed from
(6,3) to (6,2). This is followed at $t=4500$ by the reverse change where the (6,2) scheme is replaced by the (6,3) scheme.
}
\end{figure}

The snapshots in Fig. \ref{fig10} allow to gain valuable insights in how the system adjusts to a change of the interaction
scheme. The first row in the figure shows how the space-time pattern changes when switching at $t=500$ from (6,3) to (6,2).
It takes some time for this change to modify the spirals and the domains: after 20 time steps the general
spatial structure is basically unchanged, and only in a few selected spots do we see appearing small regions filled by the
new alliances. It is only later, as shown in the $t=600$ snapshot, that individuals from neutral species
start to agglomerate on a large scale. This is accompanied by the dissolution of the spirals as well as by the breaking up of the old domains, followed
by the formation of new domains that contain only members of two neutral partners forming one team. This process is rather
slow and even after 250 time steps, pockets with spirals persist. 

The second row in Fig. \ref{fig10} shows the adjustments of the system when at $t=4500$ the scheme is changed back to (6,3).
A quick, immediate, reaction is the segregation that takes place inside the domains: as the previously neutral
partners are now preying on each other (for example, species 1 and 4), we observe the immediate formation inside the larger domains of small domains that only contain
individuals from one species. Whereas these small domains coarsen, see the snapshot at $t=4600$, some first spirals are started at the interface
between the original domains. 250 time steps after the change, spirals start to be more common and first larger areas 
emerge that contain only the three teams of one of the two new alliances.

\begin{figure} [h]
\includegraphics[width=0.45\columnwidth,clip=true]{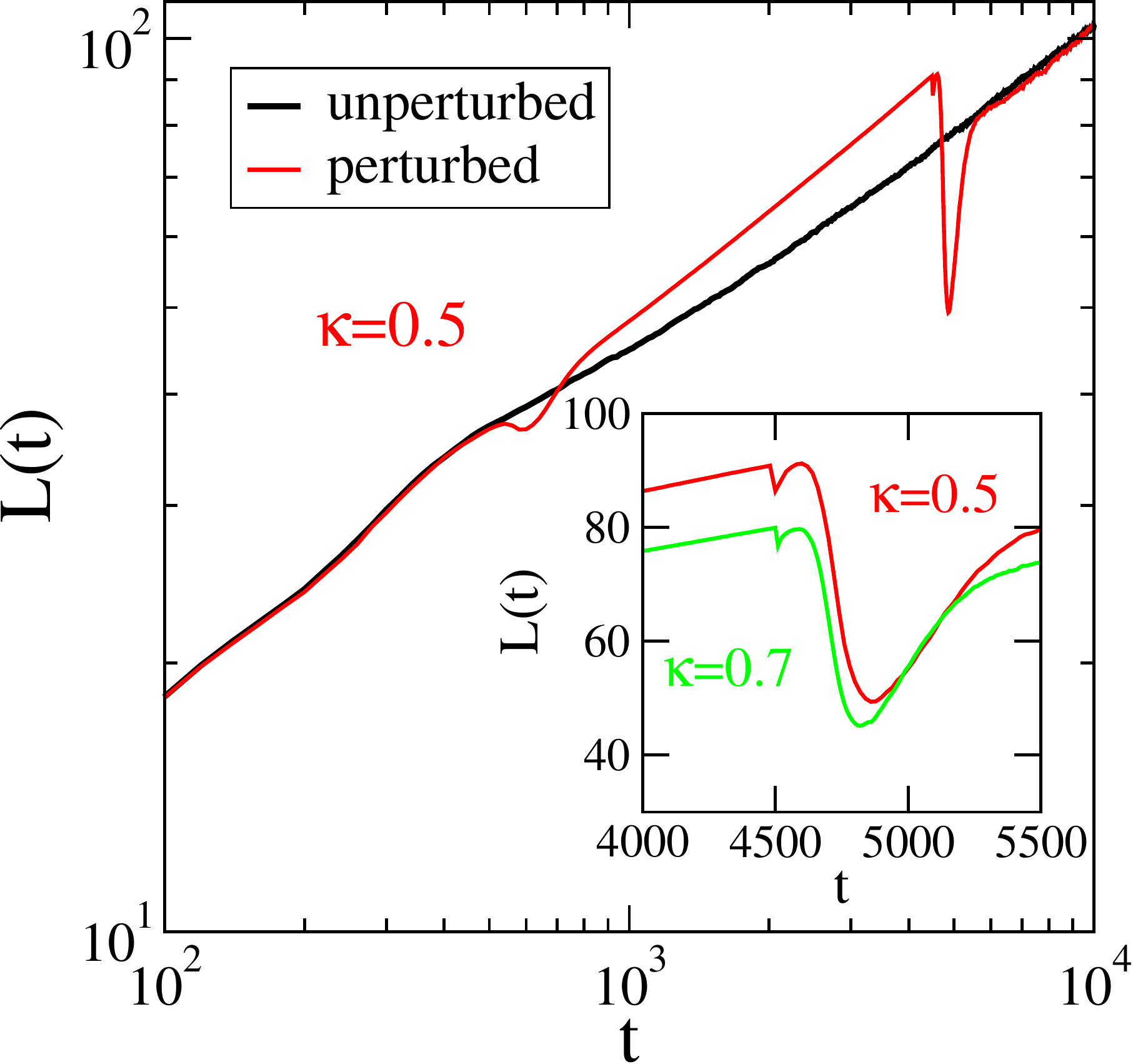}
\caption{\label{fig11} Comparison of the time-dependent correlation lengths of the unperturbed (6,3) system (thick black line)
and the system where at $t=500$ the scheme is changed from (6,3) to (6,2), followed by the change from (6,2) to (6,3) at
time $t=4500$. These curves result from the intersections of the standard space-time correlation function
with the horizontal line $C_0 = 0.05$.
The space-time correlation functions used to obtain these lengths have been averaged over
500 runs. The inset focuses on the complicated changes to the correlation length when at $t=4500$ the interaction scheme
is changed from (6,2) to (6,3) and compares the cases with $\kappa = 0.5$ and $\kappa = 0.7$. The system contains $700 \times 700$ sites and $\kappa = 0.5$.
}
\end{figure}

This discussion can be done more quantitatively by measuring the correlation length $L(t)$ from the standard
space-time correlation function (\ref{Cstandard}), see Fig. \ref{fig11}. We use a small value $C_0 = 0.05$
as this allows to obtain for both schemes a dynamical length that reveals in-domain properties as well as 
properties of the coarsening process.
We first note the dip in $L(t)$ that
takes place when we change at $t=500$ the scheme from (6,3) to (6,2). This dip reveals the dissolution of
the existing domains with three species, followed by the formation of the new domains composed by two neutral
partners. Once these new domains are formed, the standard curvature driven coarsening process of the (6,2) game
takes place, as witnessed by an increase of $L(t) \sim t^{1/2}$, see the red line. At $t=4500$, we switch back to 
the (6,3) scheme. This change results in a complicated behavior of the correlation length, see the inset in 
Fig. \ref{fig11}. The first quick decrease followed by a short increase reveals the formation inside the larger 
domains of small regions occupied by a single species and growing with time. 
The second larger drop is due to the disordering that
takes place when spirals start to form and the original domains are dissolved. Once the new domains with
in-domain spirals have been established, the coarsening process restarts and the correlation length quickly 
approaches that of a system for which the (6,3) scheme was kept at all times. 
The change in correlation length at these drops is
not very sensitive to the rates involved: for $\kappa=0.5$ resp. $\kappa = 0.7$ the distance between the preceding maximum
and the minimum of the first dip is 0.06 resp. 0.03, whereas for the larger second drop this distance is 41.87 resp. 34.77.

We also investigated systems where the perturbation has been kept for other time intervals. Qualitatively
the same features are observed in the correlation length for all studied cases, but there are some small quantitative
differences. Consider as an example the situation where the switch back to the (6,3) scheme happens at $t=6000$
instead of $t=4500$. As in that case the neutral domains that characterize the (6,2) scheme have grown for a longer time,
they are larger. Consequently, after switching back to (6,3) the disordering process is more dramatic and the second drop
in the correlation length is larger (49.72, which should be compared to 41.87 when switching the scheme at $t=4500$).
The whole process of dissolving the (6,2) domains and  forming the (6,3) domains with internal spirals also takes
slightly longer.

\begin{figure} [h]
\includegraphics[width=0.45\columnwidth,clip=true]{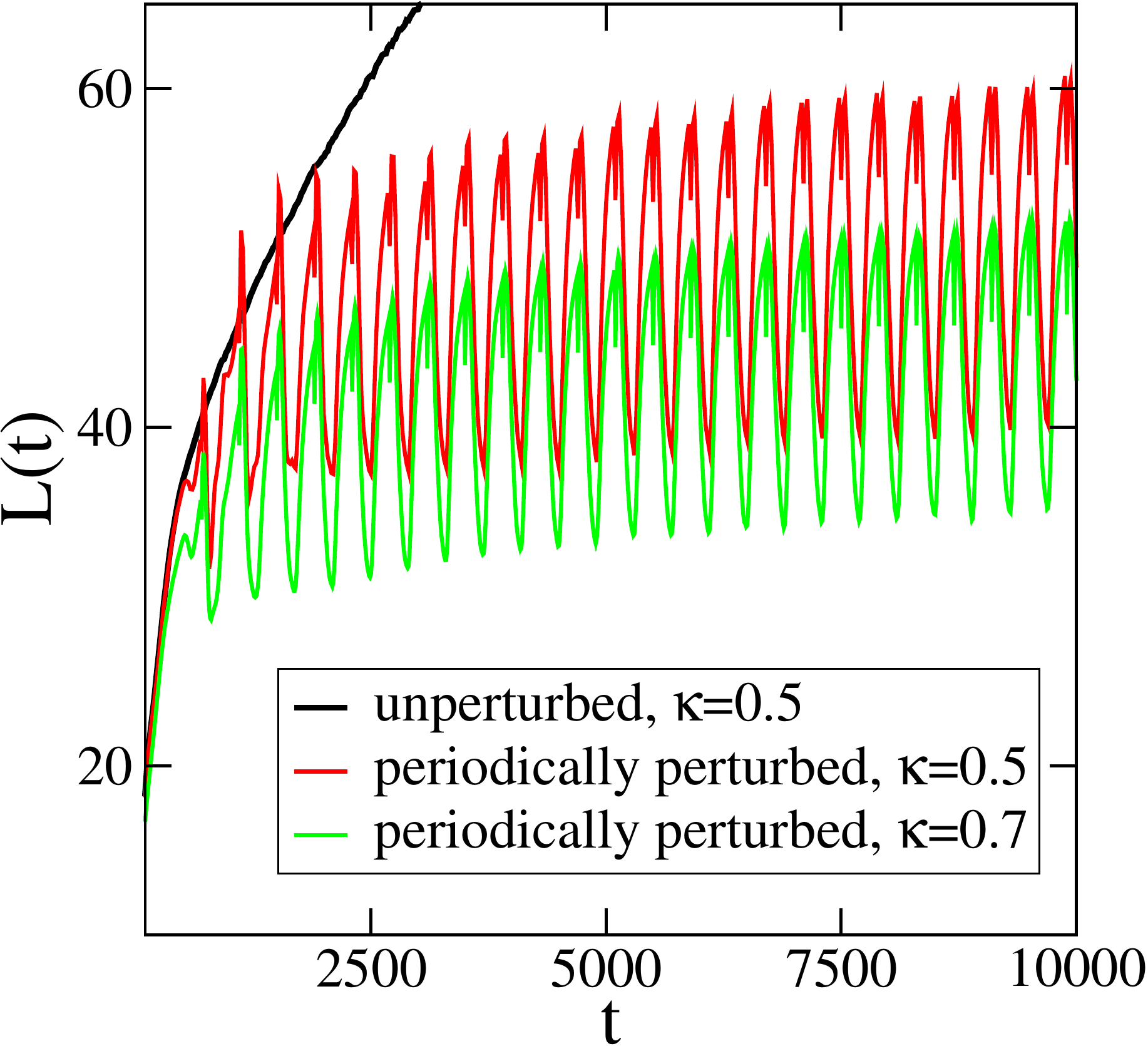}
\caption{\label{fig12} 
Time-dependent correlation lengths for systems composed of $700 \times 700$ sites with $\kappa = 0.5$
and $\kappa=0.7$ that oscillate between the (6,3) and (6,2) interaction schemes. The first change from
(6,3) to (6,2) takes place at $t=500$, followed by a change of interaction scheme every 200 time steps. The space-time
correlation function used to obtain this
length has been averaged over 500 runs, and $C_0 = 0.05$ was used.
}
\end{figure}

\begin{figure} [h]
\minipage{0.22\textwidth}
  \centering
  \includegraphics[width=\linewidth]{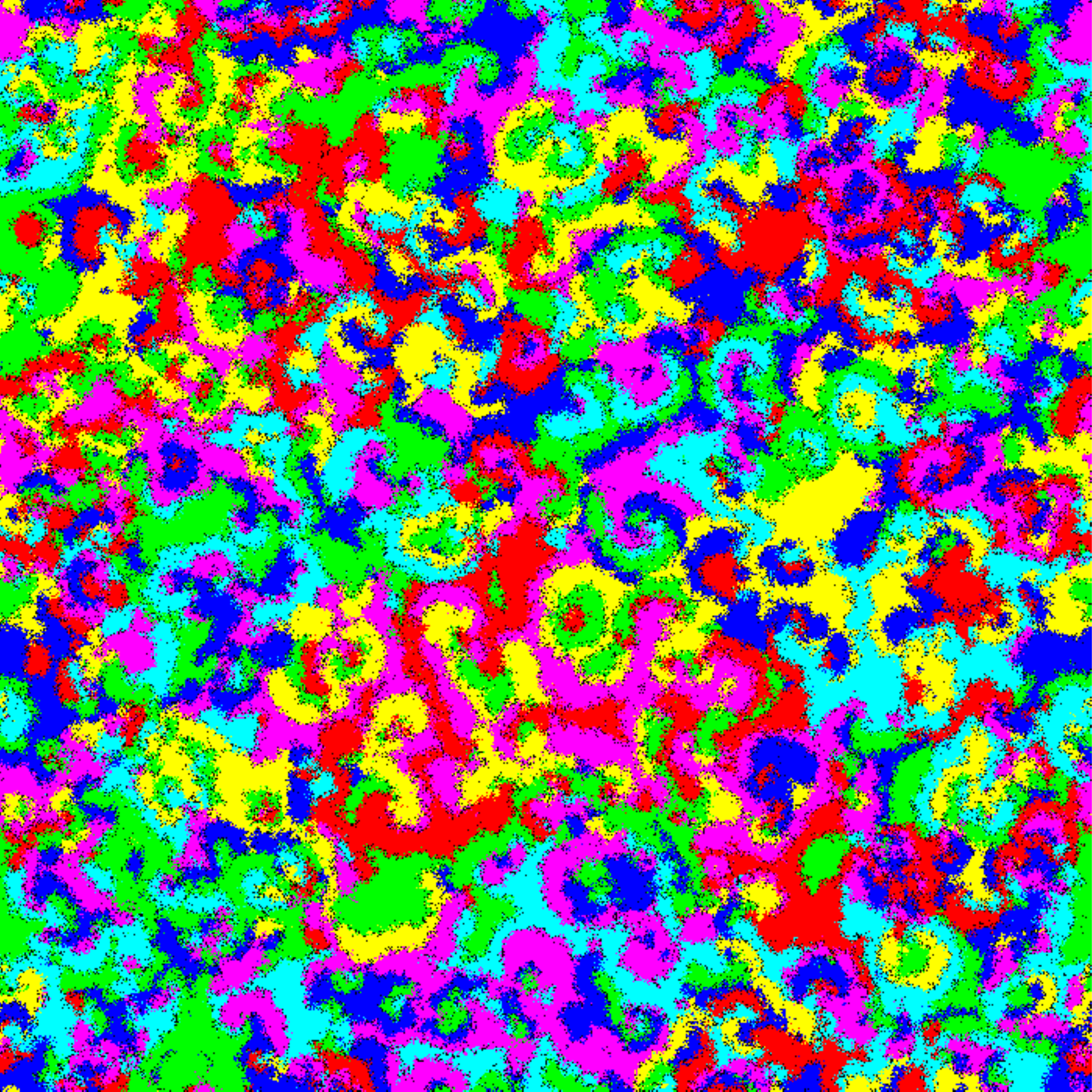}\\
  t=4900
\endminipage
\hspace*{0.03\textwidth}
\minipage{0.22\textwidth}
  \centering
  \includegraphics[width=\linewidth]{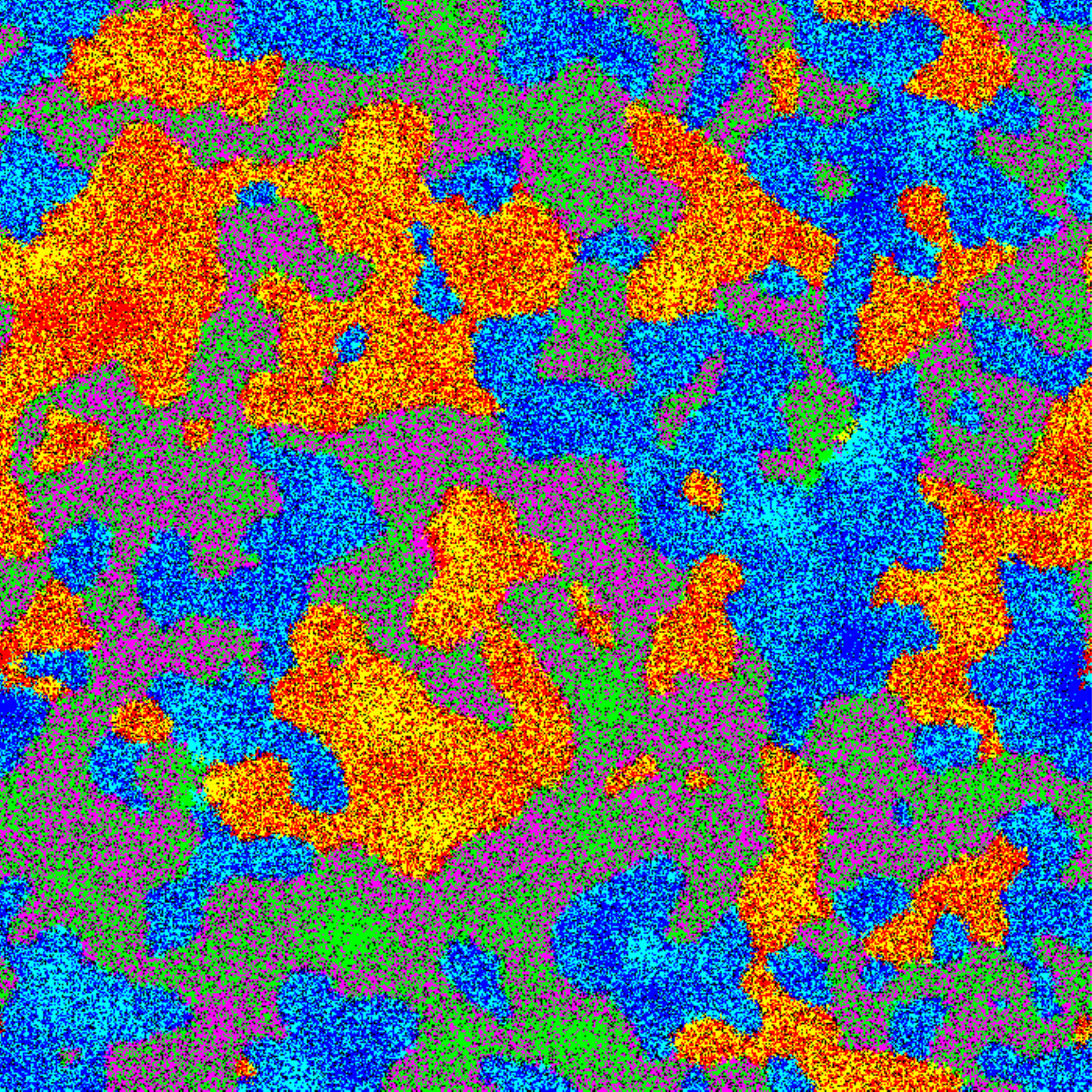}\\
  t=5100
\endminipage
\hspace*{0.03\textwidth}
\minipage{0.22\textwidth}
  \centering
  \includegraphics[width=\linewidth]{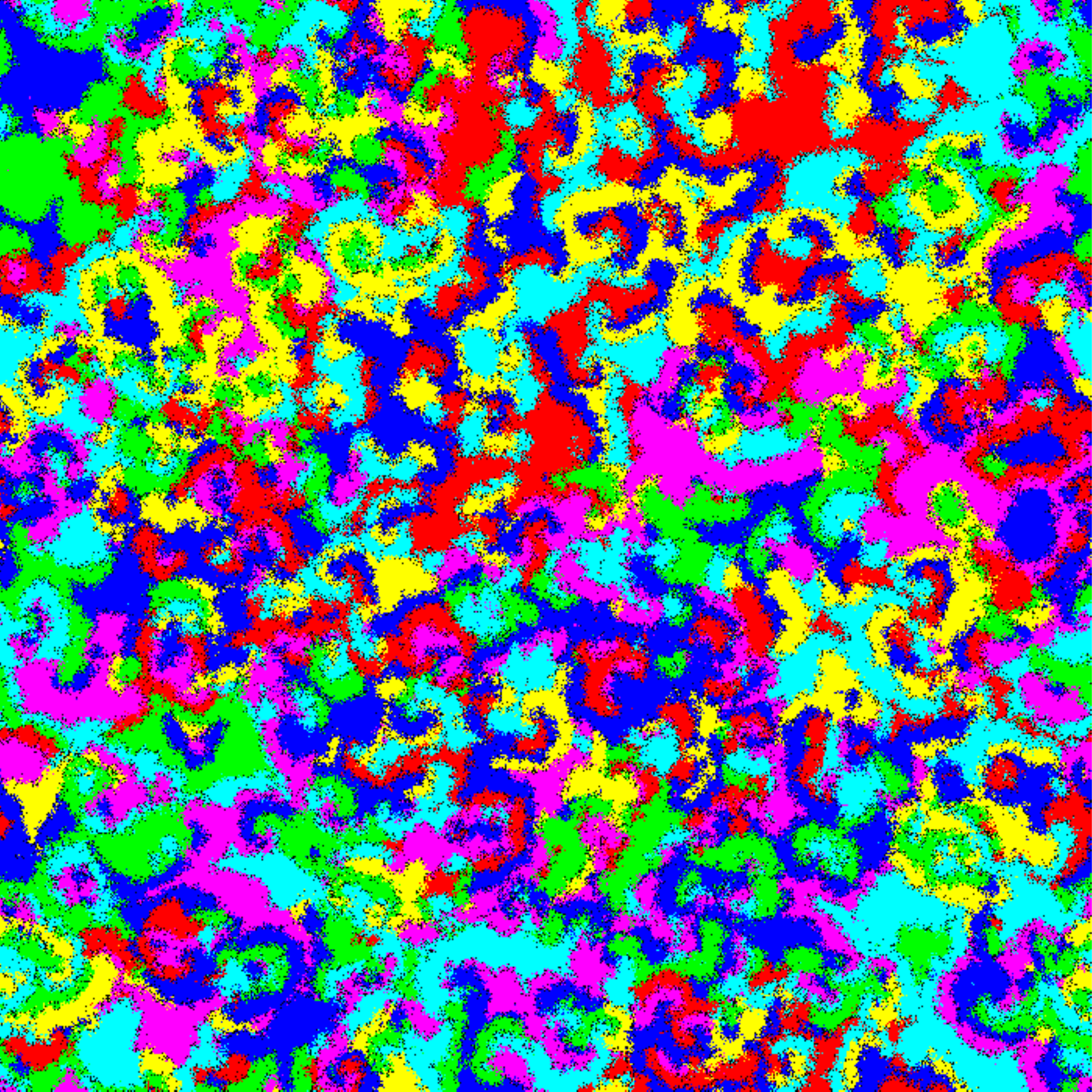}\\
  t=9700
\endminipage
\hspace*{0.03\textwidth}
\minipage{0.22\textwidth}
  \centering
  \includegraphics[width=\linewidth]{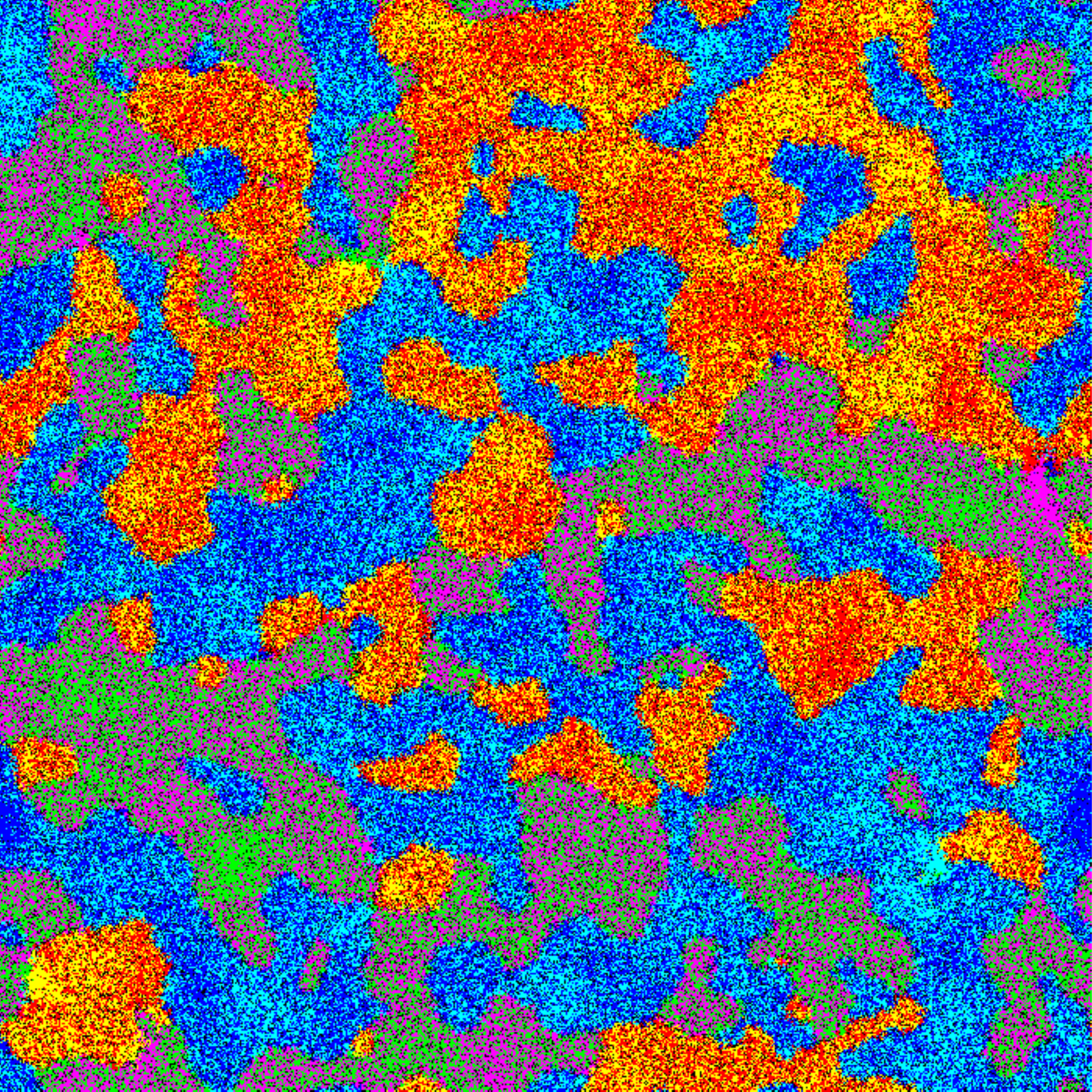}\\
  t=9900
\endminipage
\caption{\label{fig13}
Snapshots of a system composed of $700 \times 700$ sites with $\kappa = 0.5$ for which the interaction schemes changes
periodically between the (6,3) and (6,2) schemes. The first and the third panel show configurations at the moment the
scheme changes from (6,3) to (6,2), whereas the second and the forth panel show configurations at the moment the
scheme changes from (6,2) to (6,3).
}
\end{figure}

Figs. \ref{fig12} and \ref{fig13} illustrate the behavior of a system subjected to a periodic switching between
the (6,3) and (6,2) interaction schemes. Starting from a disordered initial state the system is evolved 
with the (6,3) interaction scheme until $t=500$ at which time the scheme is changed to (6,2). After this first switch
we continue switching between the two schemes after every 200 time steps. The snapshots in Fig. \ref{fig13}
reveal that the system oscillates between two states that show dominant features of either the (6,3) and (6,2) schemes,
but with some features of the inactive scheme still persisting. Based on these snapshots these hybrid states do
not change much with time. However, the correlation length displayed in Fig. \ref{fig12} indicates that the maximum length
achieved in every cycle increases slowly with time. This increase is roughly logarithmic. The periodic switching,
while being very disruptive for the ordering process, does not completely stop the growth of the dynamical length.

Fig. \ref{fig12} also provides information on the dependence of these features on the predation rate $\kappa$.
Comparison of the data for $\kappa = 0.5$ and $\kappa =0.7$ reveals that an increase of $\kappa$ yields a decrease of
the amplitude of the correlation length oscillations. Thus for $\kappa = 0.5$ the amplitude is 20.5, whereas for $\kappa=0.7$
one finds an amplitude of $17.2$. 

\section{Conclusion}
In many instances new insights into the dynamic properties of a non-equilibrium system can be gained by monitoring
the response of the system to perturbations. A well studied example is the response to switching on a  magnetic field of a 
ferromagnet initially prepared at high temperature and quenched below the critical point \cite{Henkel10}. Coarsening 
magnetic domains, however, are rather simple: the dynamics inside the domains is bulk-like and characterized by thermal 
fluctuations and the relevant degrees of freedom are provided by the interfaces separating the domains.

In this work we considered how a more complex coarsening system with non-trivial in-domain dynamics responds to
different types of perturbations. The system we have chosen is the two-dimensional (6,3) cyclic predator-prey model where each of the six species 
attack three others \cite{Roman13,Brown17}. As a result of this interaction scheme two alliances are formed that result in the competition of two types
of coarsening domains. Inside each domain the three members of an alliance perform a rock-paper-scissors game that
gives rise to the formation of spirals. This surprisingly complex system also yields complex responses when subjected to
different perturbations. We have investigated the response to changing the values of the predation and reproduction rates as well as to
changing the interaction scheme. For the latter protocol we switch during the growth process the interaction scheme from (6,3) to (6,2),
a different six-species game characterized by the competition of three types of domains containing two neutral partners. The inspection
of snapshots as well as the monitoring of time-dependent correlation functions and densities of empty sites allow to gain a
good understanding on how a system dominated by the formation of spirals inside coarsening domains
adapts to these different changes. In the three-species May-Leonard game a change of rates results in 
changes to the quantities (as for example the wavelength) characterizing the spirals \cite{Reichenbach07}. Similar quantitative
changes to the in-domain space-time patterns are observed in our study when changing the rates in the (6,3) model. Due to the competition
between the two different cyclic alliances, a modification of the spiral size impacts the coarsening process and results in an
abrupt and non-monotonous change of the correlation length. More complicated responses are encountered when changing the interaction
scheme, as the dissolution of old alliances and the formation of new ones yield a change of the in-domain dynamics as well as 
the emergence of new types of domains, followed by a different coarsening process.

Many-species predator-prey models provide many examples of intriguing space-time patterns, and some of the lessons learned in the present
study can be exploited for other situations. As an example we mention the possibility of hierarchical games with spirals within spirals \cite{Brown18}.
We also remark that the well-studied rock-paper-scissors game, a three-species game for which real-world examples are known to exist
\cite{Sinervo96,Kerr02,Kerr06,Nahum11,Kirkup04,Jackson75},
are amenable to similar perturbations (changes to the rates but also changes to the interaction scheme) that could provide a path for the
experimental realization of protocols similar to those studied in this work. We plan to investigate these and other situations in the future.

While we have assumed the special situation where each predator attacks each of their preys with the same rate, a more realistic
situation, especially in cases of environmental changes, is that of heterogeneous rates which can result in emerging cyclic alliances
that are not obvious from the definition of the model \cite{Perc07,Szabo08,Szolnoki15,Szolnoki18}. This opens the possibility for
new intriguing types of responses in many-species predator-prey systems that are not captured by the simple scenarios discussed in our work. 

To conclude, in this work we investigated the perturbation of a coarsening process with non-trivial in-domain dynamics in the form of
spirals. Whereas
our earlier work \cite{Brown17} pointed out some quantitative differences in domain growth and interface fluctuations when
comparing a standard coarsening process with a process where inside domains spirals form, the current work revealed that
perturbations of coarsening domains with spirals result in complex and novel responses that have not yet been seen in other systems
and that can not be easily anticipated from our understanding of standard coarsening processes.

\begin{acknowledgments}
This work is supported by the US National
Science Foundation through grant DMR-1606814.
\end{acknowledgments}


\begin{thebibliography}{99}

\bibitem{May74} R. M. May, {\it Stability and Complexity in Model Ecosystems} (Cambridge University Press,
Cambridge, England, 1974).
\bibitem{Smith74} J. Maynard Smith, {\it Models in Ecology} (Cambridge University Press, Cambridge, England, 1974).
\bibitem{Sole06} R. V. Sole and J. Basecompte, {\it Self-Organization in Complex Ecosystems}
(Princeton University Press, Princeton, 2006).
\bibitem{Szabo07} G. Szab\'o and G. F\'ath, Phys. Rep. {\bf 446}, 97 (2007).
\bibitem{Frey10} E. Frey, Physica A {\bf 389}, 4265 (2010).
\bibitem{McK95} A. J. McKane and T. J. Newman,
Phys. Rev. Lett. {\bf 94}, 218102 (2005).
\bibitem{Mob06} M. Mobilia, I. T. Georgiev, and U. C. T\"{a}uber,
Phys. Rev. E {\bf 73}, 040903(R) (2006).
\bibitem{Mob07} M. Mobilia, I. T. Georgiev, and U. C. T\"{a}uber,
J. Stat. Phys. {\bf 128}, 447 (2007).
\bibitem{Szolnoki14} A. Szolnoki, M. Mobilia, L. L. Jiang, B. Szczesny, A. M. Rucklidge, and M. Perc, J. Roy. Soc. Interface {\bf 11}, 20140735 (2014).
\bibitem{Dobramysl18} U. Dobramysl, M. Mobilia, M. Pleimling, and U. C. T\"{a}uber, J. Phys. A: Math. Theor. {\bf 51}, 063001 (2018). 
\bibitem{Roman13} A. Roman, D. Dasgupta, and M. Pleimling, Phys. Rev. E {\bf 87} 032148 (2013).
\bibitem{Mowlaei14} S. Mowlaei, A. Roman, and M. Pleimling, J. Phys. A: Math. Theor. {\bf 47}, 165001 (2014).
\bibitem{Roman16} A. Roman, D. Dasgupta, and M. Pleimling, J. Theor. Biol. {\bf 403}, 10 (2016).
\bibitem{Brown17} B. L. Brown and M. Pleimling, Phys. Rev. E {\bf 96}, 012147 (2017).
\bibitem{Avelino12a} P. P. Avelino, D. Bazeia, L. Losano, and J. Menezes J, Phys. Rev. E {\bf 86}, 031119 (2012).
\bibitem{Avelino12b} P. P. Avelino, D. Bazeia, L. Losano, J. Menezes, and B. F. de Oliveira,
Phys. Rev. E {\bf 86}, 036112 (2012).
\bibitem{Avelino14a} P. P. Avelino, D. Bazeia, J. Menezes, and B. F. de Oliveira, Phys. Lett. A {\bf 378}, 393 (2014).
\bibitem{Avelino14b} P. P. Avelino, D. Bazeia, J. Losano, J. Menezes, and B. F. de Oliveira,
Phys. Rev. E {\bf 89}, 042710 (2014).
\bibitem{Avelino17a} P. P. Avelino, D. Bazeia, L. Losano, J. Menezes, and B. F. de Oliveira,
Phys. Lett. A {\bf 381}, 1014 (2017).
\bibitem{Avelino17b} P. P. Avelino, D. Bazeia, L. Losano, J. Menezes, and B. F. de Oliveira,
EPL {\bf 121}, 48003 (2018).
\bibitem{Labavic16} D. Labavi\'{c} and H. Meyer-Ortmanns, J. Stat. Mech. {\bf (2016)} 113402.
\bibitem{May72} R. M. May, Nature {\bf 238}, 413 (1972).
\bibitem{Mcc00} K. McCann, Nature {\bf 405}, 228 (2000).
\bibitem{Pas06} M. Pascual and J. Dunne, {\it Ecological Networks: Linking Structure
to Dynamics in Food Webs} (Oxford University Press, Oxford, England, 2006).
\bibitem{Bas10} J. Bascompte, Science {\bf 239}, 765 (2010).
\bibitem{Henkel10} M. Henkel and M. Pleimling, {\it Non-Equilibrium Phase Transitions, Volume 2: Ageing and Dynamical
Scaling Far From Equilibrium} (Springer, Heidelberg, 2010).
\bibitem{Edwards82} S. F. Edwards and D. R. Wilkinson, Proc. R. Soc. London Ser. A {\bf 381}, 17 (1982).
\bibitem{Roman12} A. Roman, D. Konrad, and M. Pleimling, J. Stat. Mech. (2012) P07014.
\bibitem{Belleti08} F. Belletti, A. Cruz, L. A. Fernandez, A. Gordillo-Guerrero, M. Guidetti, A. Maiorano, F. Mantovani, 
E. Marinari, V. Martin-Mayor, J. Monforte, A. Muñoz Sudupe, D. Navarro, G. Parisi, S. Perez-Gaviro, J. J. Ruiz-Lorenzo, 
S. F. Schifano, D. Sciretti, A. Tarancon, R. Tripiccione, and D. Yllanes, J. Stat. Phys. {\bf 135}, 1121 (2009).
\bibitem{Park12} H. Park and M. Pleimling, Eur. Phys. J. B {\bf 85}, 300 (2012).
\bibitem{Brown15} M. O. Brown, R. H. Galyean, X. Wang, and M. Pleimling, Phys. Rev. E {\bf 91}, 052116 (2015).
\bibitem{Reichenbach07} T. Reichenbach, M. Mobilia, and E. Frey, Nature {\bf 448}, 1046 (2007).
\bibitem{Brown18} B. L. Brown, H. Meyer-Ortmanns, and M. Pleimling, in preparation.
\bibitem{Sinervo96} B. Sinervo and C. M. Lively, Nature {\bf 380}, 240 (1996).
\bibitem{Kerr02} B. Kerr, M. A. Riley, M. W. Feldman, and B. J. M. Bohannan, Nature {\bf 418}, 171 (2002).
\bibitem{Kerr06} B. Kerr, C. Neuhauser, B. J. M. Bohannan, and A. M. Dean, Nature {\bf 442}, 75 (2006).
\bibitem{Nahum11} J. R. Nohum, B. N. Harding, and B. Kerr, Proc. Natl. Acad. Sci. {\bf 108}, 10831 (2011).
\bibitem{Kirkup04} B. C. Kirkup and M. A. Riley, Nature {\bf 428}, 412 (2004).
\bibitem{Jackson75} J. B. C. Jackson and L. Buss, Proc. Nat. Acad. Sci. {\bf 72}, 5160 (1975).
\bibitem{Perc07} M. Perc, A. Szolnoki, and G. Szab\'{o}, Phys. Rev. E {\bf 75}, 052102 (2007).
\bibitem{Szabo08} G. Szab\'{o}, A. Szolnoki, and I. Borsos, Phys. Rev. E {\bf 77}, 041919 (2008).
\bibitem{Szolnoki15} A. Szolnoki and M. Perc, EPL {\bf 110}, 38003 (2015).
\bibitem{Szolnoki18} A. Szolnoki and M. Perc, New. J. Phys. {\bf 20}, 013031 (2018).

\end{thebibliography}
\end{document}